\documentclass[a4paper,fleqn,usenatbib]{mnras}
\usepackage[T1]{fontenc}
\usepackage{ae,aecompl}
\usepackage{graphicx}
\usepackage{mathrsfs}
\usepackage{amsmath}
\usepackage{amssymb}
\usepackage{verbatim}
\def\apjl{ApJ}
\def\araa{ARA\&A}             
\def\apj{ApJ}                 
\def\apjl{ApJ}                
\def\apjs{ApJS}               
\def\ao{Appl.~Opt.}           
\def\aap{A\&A}                
\def\aaps{A\&AS}              
\def\mnras{MNRAS}             
\def\qjras{QJRAS}             
\def\nat{Nature}              

\def\memsai{Mem.~Soc.~Astron.~Italiana}

\title[Timing of the AMXP SAX J1748.9-2021]{Timing of the accreting millisecond pulsar SAX J1748.9$-$2021 during its 2015 outburst}
\author[Sanna et al. ]{A. Sanna$^{1}$\thanks{E-mail:
    andrea.sanna@dsf.unica.it}, L. Burderi$^{1}$, A. Riggio$^{1}$, F. Pintore$^{2,1}$,T. Di Salvo$^{3}$, A. F. Gambino$^{3}$ \newauthor
      R. Iaria$^{3}$, M. Matranga$^{3}$, F. Scarano$^{1}$\\
$^{1}$Dipartimento di Fisica, Universit\`a degli Studi di Cagliari, SP Monserrato-Sestu km 0.7, 09042 Monserrato, Italy\\
$^{2}$INAF-Istituto di Astrofisica Spaziale e Fisica Cosmica - Milano, via E. Bassini 15, I-20133 Milano, Italy\\
$^{3}$Universit\`a degli Studi di Palermo, Dipartimento di Fisica e Chimica, via Archirafi 36, 90123 Palermo, Italy}
\begin{document}

\date{Accepted -. Received -; in original form -}

\pagerange{\pageref{firstpage}$-$\pageref{lastpage}} \pubyear{2015}

\maketitle

\label{firstpage}

\begin{abstract}
We report on the timing analysis of the 2015 outburst of the intermittent accreting millisecond X-ray pulsar SAX J1748.9$-$2021 observed on March 4 by the X-ray satellite \textit{XMM-Newton}. By phase-connecting the time of arrivals of the observed pulses, we derived the best-fit orbital solution for the 2015 outburst. We investigated the energy pulse profile dependence finding that the pulse fractional amplitude increases with energy while no significant time lags are detected. Moreover, we investigated the previous outbursts from this source, finding previously undetected pulsations in some intervals during the 2010 outburst of the source. Comparing the updated set of orbital parameters, in particular the value of the time of passage from the ascending node, with the orbital solutions reported from the previous outbursts, we estimated for the first time the orbital period derivative corresponding with $\dot{P}_{orb}=(1.1\pm0.3)\times 10^{-10}$ s/s. We note that this value is significant at $3.5\sigma$ confidence level, because of significant fluctuations with respect to the parabolic trend and more observations are needed in order to confirm the finding. Assuming the reliability of the result, we suggest that the large value of the orbital-period derivative can be explained as a result of an highly non-conservative mass transfer driven by emission of gravitational waves, which implies the ejection of matter from a region close to the inner Lagrangian point. We also discuss possible alternative explanations.
\end{abstract}


\section{Introduction}
Accretion-powered millisecond X-ray Pulsars (AMXPs) are transient low-mass X-ray binaries (LMXBs) showing X-ray pulsations during the outburst phases at frequencies larger than $\sim$100 Hz \citep{Alpar82}. Matter transferred from
the companion star via Roche-lobe overflow is captured by the neutron star (NS) magnetosphere and forced to follow 
the magnetic lines down to the NS's magnetic polar caps. Among the 18 known AMXPs \citep{Burderi13, Patruno12b, Papitto2015a}, fifteen show 
persistent X-ray pulsations throughout the outbursts \citep[with PSR J1023+0038
 and XSS J12270 showing persistent pulsations at a much lower luminosity than those of the canonical AMXPs;][]{Archibald2015a, Papitto2015a}. The three remaining sources only occasionally show X-ray 
pulsations: Aql X-1 \citep{Casella08} showed pulsations only during a 150 s segment of data over more than 
1.3 Ms available, HETE J1900.1$-$2455 \citep{Kaaret06} switched off the X-ray pulsations after 2 month from the 
beginning of a long outburst, and SAX J1748.9$-$2021 \citep{Gavriil07, Altamirano08a,Patruno09a} for which the X-ray 
pulsations turned on and off intermittently during the outbursts. What makes these 3 sources different from the rest 
of the known AMXPs is still unclear. However, solving this issue could help to understand the lack of pulsations in a large number 
of LMXBs (around 100 at the moment).

SAX J1748.9$-$2021 is a NS X-ray transient hosted in the globular cluster NGC 6440 located at 8.5$\pm$0.4 kpc \citep{Ortolani94}. The source was discovered by BeppoSax in 1998 during monitoring of the X-ray activity around the Galactic center \citep{in-t-Zand99}. Since then, SAX J1748.9$-$2021 has been observed in outburst 4 more times: 2001 \citep{in-t-Zand01}, 2005 \citep{Markwardt05}, 2010 \citep{Patruno10a} and recently at the beginning of 2015 \citep{Bozzo15}. X-ray pulsations at the frequency of $\sim$442.3 Hz were discovered for the first time in a single observation of the 2005 outburst \citep{Gavriil07}. More observations with pulsations have been found later on by re-analysing archival data. A first estimation of the spin frequency and the orbital parameters of SAX J1748.9$-$2021 have been reported by \cite{Altamirano08a} analysing the 2001 outburst. Using the same set of data, but applying a phase-coherent timing technique, \cite{Patruno09a} managed to determine a refined timing solution (see Tab.~\ref{tab:solution}). According to \cite{Altamirano08a}, the companion star might be a main-sequence (or a slightly evolved) star with mass ranging between 0.85 M$_\odot$ and 1.1 M$_\odot$.

Here we present the analysis of the timing properties of the coherent signal emitted by the intermittent source SAX J1748.9$-$2021, using the \textit{XMM-Newton} observation performed during the latest outburst. Moreover, we investigate the orbital evolution of the source by means of the modelling of the times of the ascending nodes determined during different outbursts.
\section[]{Observations and Data analysis}

\subsection{XMM-Newton}
\label{sec:XMM}
\begin{figure}
\centering
\includegraphics[width=0.5\textwidth]{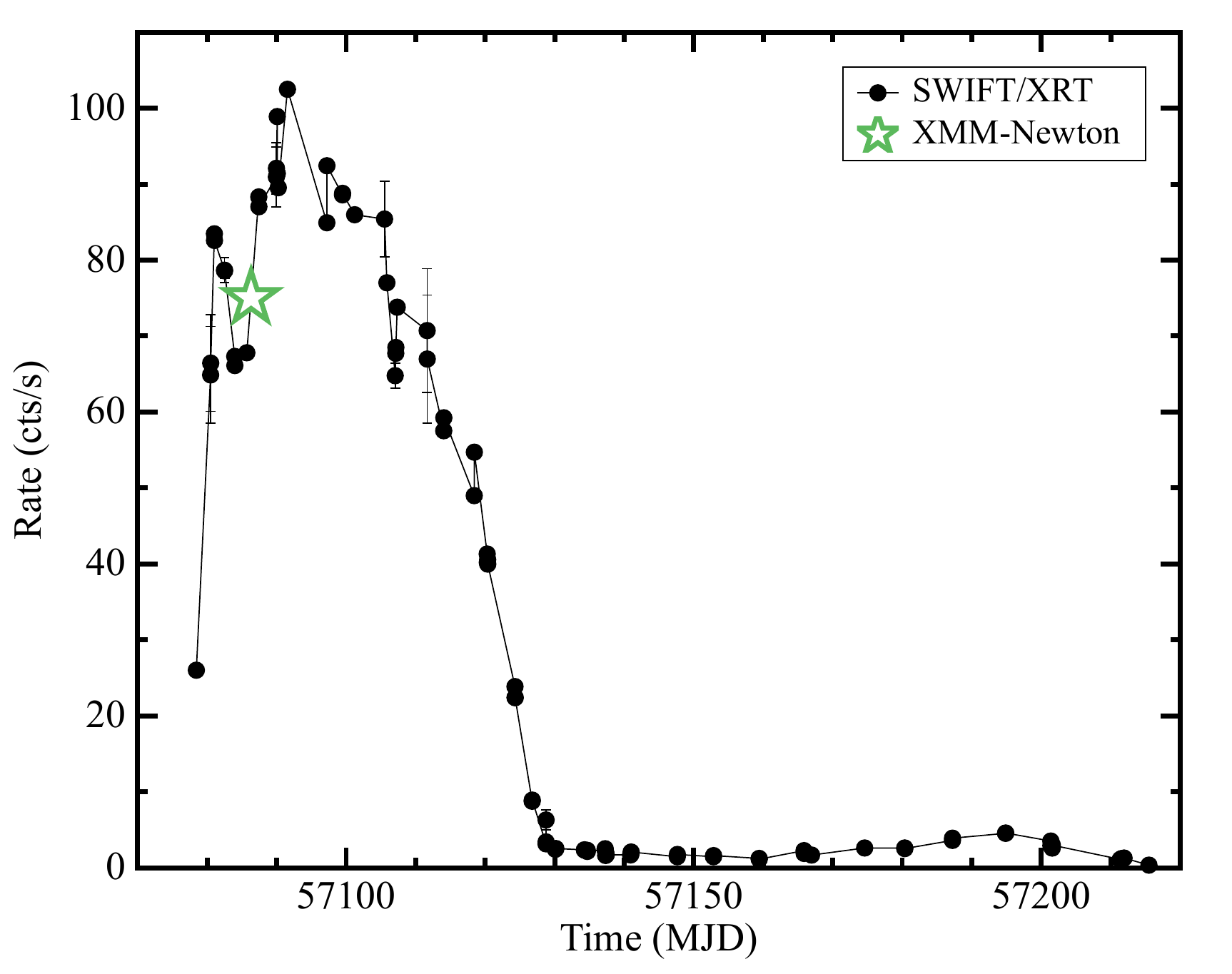}
\caption{Light-curve of the 2015 outburst of SAX J1748.9$-$2021 as observed by \textit{Swift}-XRT (black points). The green star represents the epoch of the \textit{XMM-Newton} observation}.
\label{fig:lc}
\end{figure}

We reduced the pointed \textit{XMM-Newton} observation of SAX J1748.9$-$2021 performed on March the 4th, 2015 (Obs ID 0748391301). The observation was
taken in timing mode for $\sim100$ ks and in burst mode for $\sim10$ ks, for a total exposure time of $\sim110$ ks. The combination of short exposure time and the low number of detected photons made the observation taken in burst mode not suitable for this work. We therefore decided to exclude it from the analysis. From here on with EPIC-pn (PN) data we will refer to the observation segment performed in timing mode. Fig.~\ref{fig:lc} shows the light curve of the 2015 outburst of the source monitored by \textit{Swift}-XRT (black points). The green star represents the \textit{XMM-Newton} data taken roughly a weak before the outburst peak. For this analysis we focused on the PN data, which have both the statistics and time resolution (30 $\mu$s) required to investigate the millisecond variability of the source. We performed the reduction of the PN data using the Science Analysis Software (SAS) v. 14.0.0 with the up-to-date calibration files, and adopting the standard reduction pipeline RDPHA \citep[see][for more details on the method]{Pintore14}. We verified that no significant high background flaring activity was present during the observation. We filtered the PN data in the energy range 0.3$-$15.0 keV, selecting events with \textsc{pattern$\leq$4} allowing for single and double pixel events only. \\
The PN average count rate during the observation was $\sim700$ cts/s. We estimated the background mean count rate in the RAWX range [3:5] to be $\sim1.5$ cts/s in the energy range 0.3$-$15.0 keV. During the observation numerous type-I burst episodes have been recorded, the episodes occurred almost regularly every hour, with an average duration of $\sim100$ seconds \citep[see][for a detailed analysis]{Pintore2016a}. We did not exclude the X-ray type-I burst for the timing analysis (see Sec.~\ref{sec:ta2015} for more details).

\subsection{Rossi X-ray Timing Explorer}
With the aim of improving the orbital solution for SAX J1748.9$-$2021 we also re-analysed the previous four outbursts observed by \textit{Rossi X-ray Timing Explorer} (RXTE), see Tab.~\ref{tab:obs}. In particular, we extracted data from the proportional counter array \citep[PCA; see][]{Jahoda06} instrument on board on the \textit{RXTE} satellite. We used data taken by the PCA in Event (122 $\mu$s temporal resolution) and Good Xenon (1$\mu$s temporal resolution) packing modes. Following \citet{Patruno09a}, to improve the chances to detect the X-ray pulsations, we selected the event files in the energy range between 5 and 25 keV. This energy selection allows to avoid strong background contamination at high energies and to exclude energy intervals where the pulsed fraction is below $\sim1$\% rms, helping to maximise the signal-to-noise ratio. 
\begin{table}
    \caption{Observations analysed for each outburst.}
    \label{tab:obs}
    \resizebox{0.47\textwidth}{!}{
    \begin{tabular}{c|c|c|c|c|c}
    	\hline
	\hline
      Outburst & Begin & End & Instr&Exp & Obs ID \\
      (year) & (MJD) & (MJD) & & (ks)&\\
      
      1998 & 51051.28 & 51051.45 & RXTE & 15 & P30425 \\
     
      2001 & 52138.40 & 52198.31 & RXTE & 115 & P60035/84\\
     
      2005 & 53514.30 & 53564.85 & RXTE & 19 & P91050\\
      
      2010 & 55214.79 & 55254.56& RXTE &  216 & P94315 \\
      
      2015 &57085.74 & 57086.89& XMM & 100 & 0748391301\\
      \hline
      \hline
    \end{tabular}
    }
\end{table}

\subsection{Solar-system barycentric corrections}
We corrected the PN and \textit{RXTE} photon arrival times for the motion of the Earth-spacecraft system with respect to the Solar System barycentre (DE-405 Solar System ephemeris) by using the \textsc{barycen} and the \textsc{faxbary} tools, respectively. We used the best available source position obtained with \textit{Chandra} reported by \citet[][]{Pooley02}, and reported in Tab.~\ref{tab:solution}. Using the expression of the residuals induced by the motion of the Earth for small variations of the source position $\delta_{\lambda}$ and $\delta_{\beta}$ expressed in ecliptic coordinates $\lambda$ and $\beta$ \citep[see, e.g.,][]{Lyne90}, we estimated the systematic uncertainties induced by the source position uncertainties on the linear and quadratic terms of the pulse phase delays, which correspond to an additional error in the spin frequency correction and in the spin frequency derivative, respectively. The former and the latter terms can be expressed as $\sigma_{\nu_{pos}}\leq \nu_0\,y\,\sigma_{\gamma}(1+\sin^2\beta)^{1/2}2\pi/P_{\oplus}$ and $\sigma_{\dot{\nu}_{pos}}\leq \nu_0y\sigma_{\gamma}(1+\sin^2\beta)^{1/2}(2\pi/P_{\oplus})^2$, respectively, where $y=r_E/c$ is the semi-major axis of the orbit of the Earth in light-seconds, $P_{\oplus}$ is the Earth orbital period, and $\sigma_{\gamma}$ is the positional error circle. Considering the positional uncertainty of $0.6 ''$ reported by \citet{Pooley02}, we estimated $\sigma_{\nu_{pos}} \leq 6\times 10^{-10}$ Hz and $\sigma_{\dot{\nu}_{pos}}\leq 1.4\times 10^{-16}$ Hz s$-1$. The level of accuracy of the source position guarantees us sufficient precision to proceed with a phase-coherent timing analysis of the data. These systematic uncertainties will be added in quadrature to the statistical errors estimated from the timing analysis. 

\subsection{Timing analysis of the 2015 outburst}
\label{sec:ta2015}
Starting from the timing solution inferred by \citet[][see Tab.~\ref{tab:solution}; hereafter P09]{Patruno09a} during the 2001 outburst, we corrected all the photon time of arrivals of the PN dataset for the delays caused by the binary motion applying the orbital parameters through the recursive formula  
\begin{eqnarray}
\label{eq:barygen} 
t + \frac{z(t)}{c} = t_{arr},
\end{eqnarray}
where $t$ is photon emission time, $t_{arr}$ is the photon arrival time to the Solar System barycentre, $z(t)$ is the projection along the line of sight of the distance between the NS and the barycenter of the binary system, and $c$ is the speed of light. As reported by \citet{Burderi07}, for almost circular orbits (eccentricity $e \ll 1$) we have:
\begin{eqnarray} 
\label{eq:bary}
\frac{z(t)}{c}= \frac{a \sin i}{c}\,\sin\Big(\frac{2\pi}{P_{orb}} \,(t-T^\star)\Big),
\end{eqnarray}
where $a \sin{\textit{i}/c}$ is the projected semimajor axis of the NS orbit in light seconds, $P_{orb}$ is the orbital period, and $T^\star$ is the time of passage from the ascending node.
The correct emission times (up to an overall constant $D/c$ , where $D$ is the distance between the Solar System barycenter and the barycenter of the binary system) are calculated by solving iteratively the aforementioned equation~(\ref{eq:barygen}), $t_{n+1} = t_{arr} - z(t_{n})/c$,
with $z(t)/c$ defined as in equation~(\ref{eq:bary}),
with the conditions $D/c = 0$, and $z(t_{n=0}) = 0$. We iterated until the difference between two consecutive steps (${\Delta t}_{n+1} = t_{n+1} - t_{n}$) is of the order of the absolute timing accuracy of the instrument used for the observations. In our case we set ${\Delta t}_{n+1}=1 \mu$s.

To look for pulsations we performed an epoch-folding search of the whole observation using 16 phase bins and starting with the spin frequency value $\nu_0$ = 442.36108118 Hz, corresponding to the spin frequency measured from the 2001 outburst (the most accurate spin estimate reported in literature). Given the poor knowledge of the NS spin evolution between the outbursts under consideration, we explored the frequency space around $\nu_0$ with steps of $10^{-8}$ Hz for a total of 1001 steps. We found no evidence for X-ray pulsation in the observation. 

\begin{figure}
\centering
\includegraphics[width=0.47\textwidth]{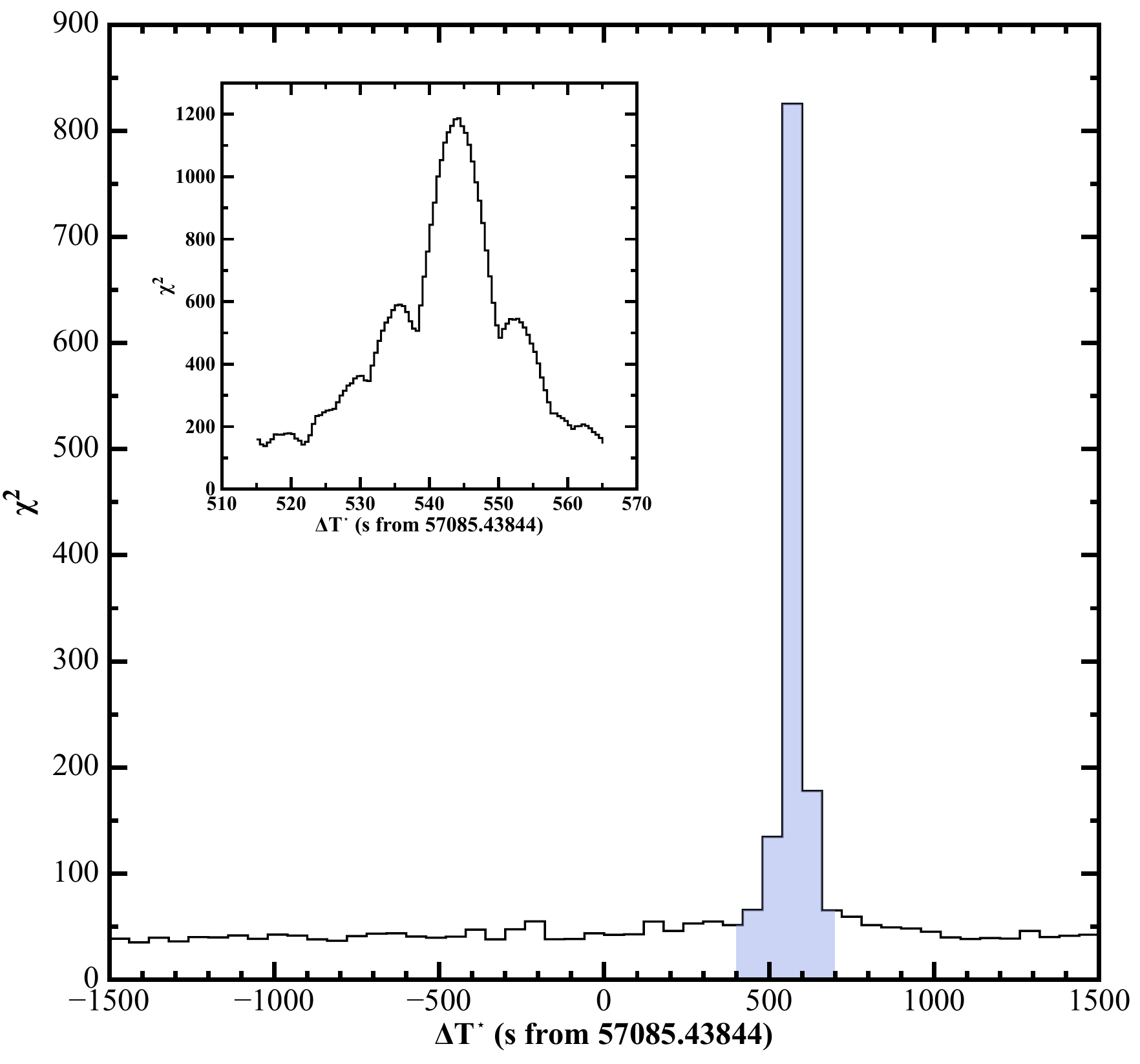}
\caption{Maximum value of $\chi^2$ from the epoch-folding search on the PN data as a function of the $T^{\star}$ values used to correct for the orbital modulation in the range $T_{2015}^{\star}\pm\sigma_{ T^{\star}_{2015}}$. $\Delta T^{\star}$ represents the delay in seconds from the predicted $T^{\star}_{2015} = 57085.43844$ MJD extrapolated from the timing solution obtained analysing the 2001 outburst. The inset shows a more detailed investigation of the region around the $\chi^2$ peak (shaded region) using a $\Delta T^{\star}$ step of 0.5s.  }
\label{fig:tstar_2015}
\end{figure}
For this reason we investigated the possibility that the lack of pulsation reflects a wrong set of orbital parameters for the source. The ephemerides of the source are expected to vary with time following the system evolution. However, the accuracy of the X-ray timing solutions for AMXPs is such that we cannot usually track variations of parameters such as the orbital period and the projected semimajor axis of the NS orbit in between outbursts. On the other hand we are often sensitive to variations of the time of passage from the ascending node \citep[e.g.,][]{Riggio11b}. From the timing solution reported by P09 we noted that $T^{\star}$ is the parameter with the largest uncertainty, and propagating the error to the 2015 outburst we found:
\begin{eqnarray}
\label{eq:sigmaTstar} 
\sigma_{ T^{\star}_{2015}} = (\sigma^2_{ T^{\star}_{2001}} + N^2 \times \sigma^2_{P_{orb_{2001}}})^{1/2}\simeq1340\, s,
\end{eqnarray}
where, $\sigma_{ T^{\star}_{2015}}$ is the error on $T^{\star}$ extrapolated from the 2001 timing solution, $\sigma_{ T^{\star}_{2001}}$ and $\sigma_{P_{orb_{2001}}}$ are the 1$\sigma$ errors on $T^{\star}$ and $P_{orb}$ from the 2001 outburst reported by P09 (see Tab.~\ref{tab:solution}), respectively, and N is the number of integer orbital cycles completed in the time interval between the outbursts. We note that this estimation has been made assuming a zero orbital-period derivative $\dot{P}_{orb}$. 

Following \citet[][see also \citealt{Riggio11b}]{Papitto05}, we investigated the orbital solution, under the assumption that the best set of orbital parameters is the one for which the folded pulse profile obtained by epoch-folding the data has the highest signal-to-noise ratio, hence the largest $\chi^2$ value in an epoch-folding search \citep[see e.g.][]{Kirsch04}. Here we focused on the $T^{\star}$ because is the parameter with largest uncertainty among the orbital parameters. We explored possible values of the parameter in the interval $T_{2015}^{\star}\pm\sigma_{ T^{\star}_{2015}}$. We corrected each time series with Eq.~\ref{eq:bary} adopting the same orbital parameters, except for $T^{\star}$, which varied in steps of 60 seconds. We then applied the epoch-folding techniques to search for X-ray pulsation around the spin frequency $\nu_0$ using 16 phase bin to sample the signal. In  Fig.~\ref{fig:tstar_2015} we report the largest value of $\chi^2$ from the epoch-folding search of each time series as a function of the $T^{\star}$ value used to correct the photon times of arrival. A clear peak is present at $\Delta T^{\star}\simeq 540$ seconds. As shown in the inset of Fig.~\ref{fig:tstar_2015}, adopting a finer stepping in $T^{\star}$ (0.5 seconds) around the value which gave the highest $\chi^2$, we were able to refine the measurement of the parameter. By fitting the top of the $\chi^2$ curve with a Gaussian plus a constant we obtained a value of $\Delta T^{\star}=543.7$ seconds. As described by \citet{Riggio11b}, the folding search technique used for this analysis does not provide a straightforward method to estimate the uncertainty on the derived $T^{\star}$ parameter. Following \citet{Riggio11b}, we then performed Monte Carlo simulations generating 100 datasets (allowing $1\sigma$ error estimations) with the same properties of the real data such as, length, count rate, pulsation fractional amplitude and orbital modulation. Applying the method previously described we derived a value of $T^{\star}$ for each simulated dataset. We defined the $1\sigma$ error interval of the time of passage from the ascending node as the standard deviation of the $T^{\star}$ distribution from the simulation. Therefore, we derived $T^{\star}$ during the 2015 outburst as $T_{2015}^{\star} = T_{2001}^{\star} + \Delta T^{\star} = 57085.444732(2)$ MJD(TDB). 

\begin{figure}
\centering
\includegraphics[width=0.47\textwidth]{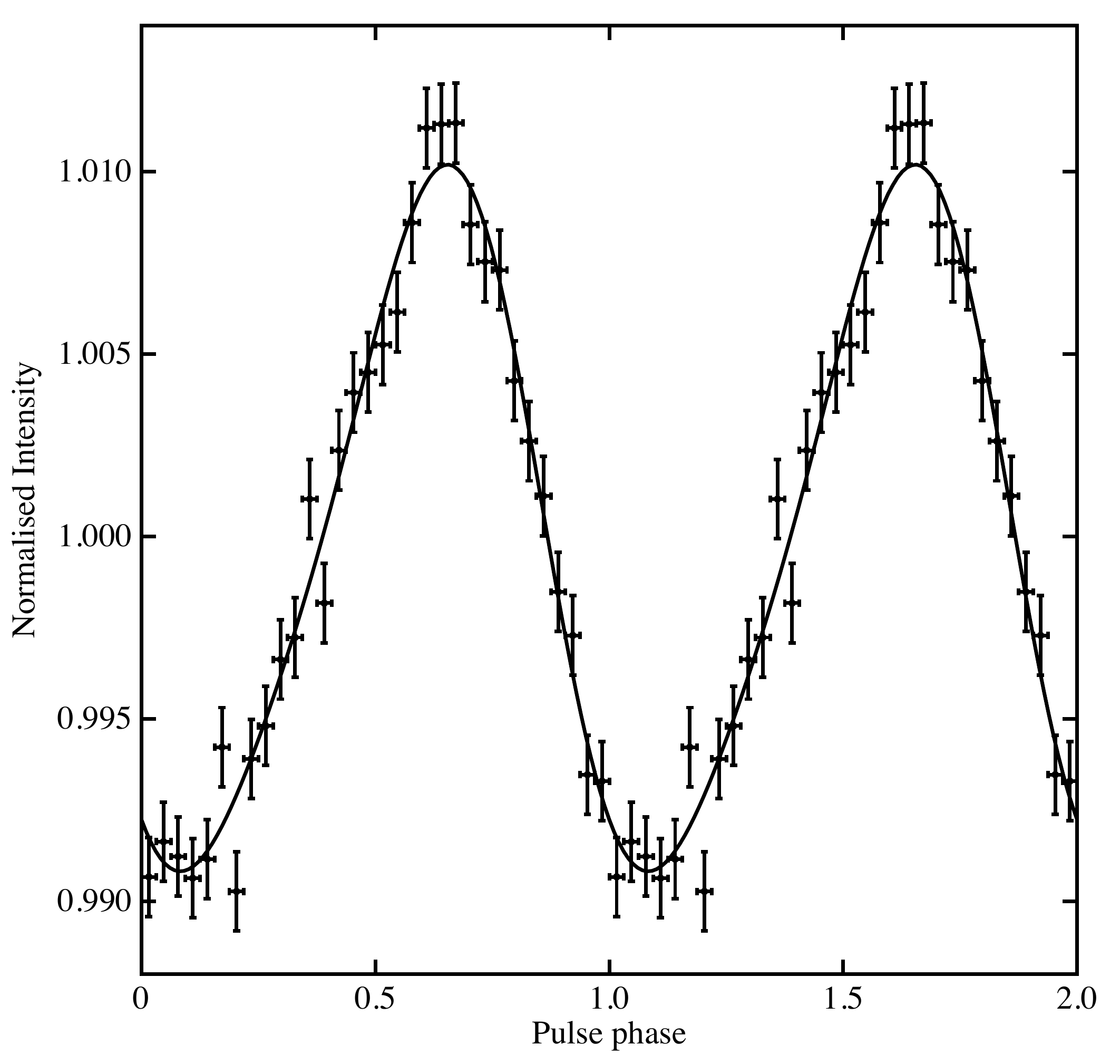}
\caption{Pulse profile and best-fitting model (combination of two sinusoids) obtained by epoch-folding the PN observation. The profile has been created after subtracting the background and it is normalised to the average flux. For clarity, we show two cycles of the pulse profile.}
\label{fig:pp}
\end{figure}

Using the updated set of orbital parameters we barycentered the PN data and performed an epoch-folding search to estimate an average local spin frequency, finding the value $\bar{\nu}=442.3610955(5)$ Hz. The error on the spin frequency has been estimated by means of Monte Carlo simulations following the method mentioned above. In Fig.~\ref{fig:pp} we show the folded pulse profile obtained epoch-folding the PN observation at $\bar{\nu}$ and sampling the signal in 32 phase bins. The pulse shape is well fitted with a combination of two sinusoids, where the fundamental and its first overtone have fractional amplitude of 0.9\% and 0.1\%, respectively.

The same analysis has been also done excluding the X-ray bursts from the data. We did not find any significant variation in terms of detectability of the pulse profile or in terms of pulse fractional amplitude. Furthermore, we investigated the presence of coherent pulsation during each of the X-ray bursts, but we found no significant evidence. We decided to continue the timing analysis including the X-ray bursts.

In order to compute statically significant pulse profiles in time intervals shorter then the whole PN observation, we split the data in time intervals of approximately 4000 seconds that we epoch-folded in 16 phase bins at the mean spin frequency $\bar{\nu}$ with respect to the epoch $T_0=57085.7$ MJD. We modelled each epoch-folded pulse profile with a sinusoid of unitary period in order to determine the corresponding sinusoidal amplitude and the fractional part of phase residual. Only folded profiles with ratio between sinusoidal amplitude and $1\sigma$ error larger than 3 were taken into consideration. We detected pulsation in $\sim81$ks of data out of the total $\sim95$ks analysed, corresponding to $\sim85\%$ of the PN observation. The fractional amplitude of the signal varies between $\sim0.6\%$ and $\sim1.6\%$, with a mean value of $\sim1\%$. 

\begin{figure}
\centering
\includegraphics[width=0.48\textwidth]{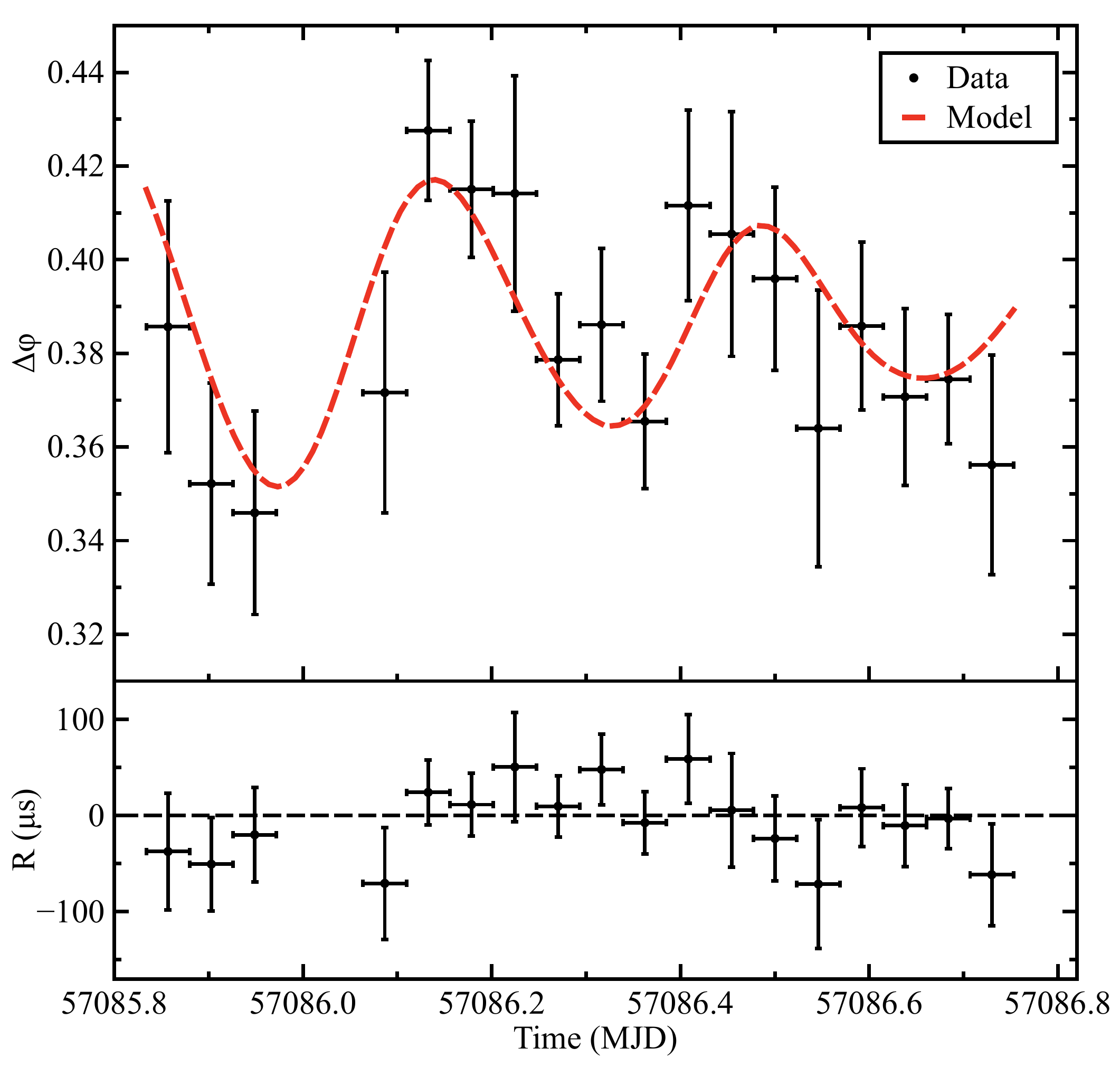}
\caption{\textit{Top panel -} Pulse phase delays as a function of time computed by epoch-folding the \textit{XMM-Newton} observation at the spin frequency $\nu_0 = 442.3610955$ Hz, together with the best-fit model (red dotted line, see text). \textit{Bottom panel -} Residuals in $\mu s$ with respect to the best-fitting orbital solution.}
\label{fig:phase_fit}
\end{figure}

\begin{figure}
\centering
\includegraphics[width=0.5\textwidth]{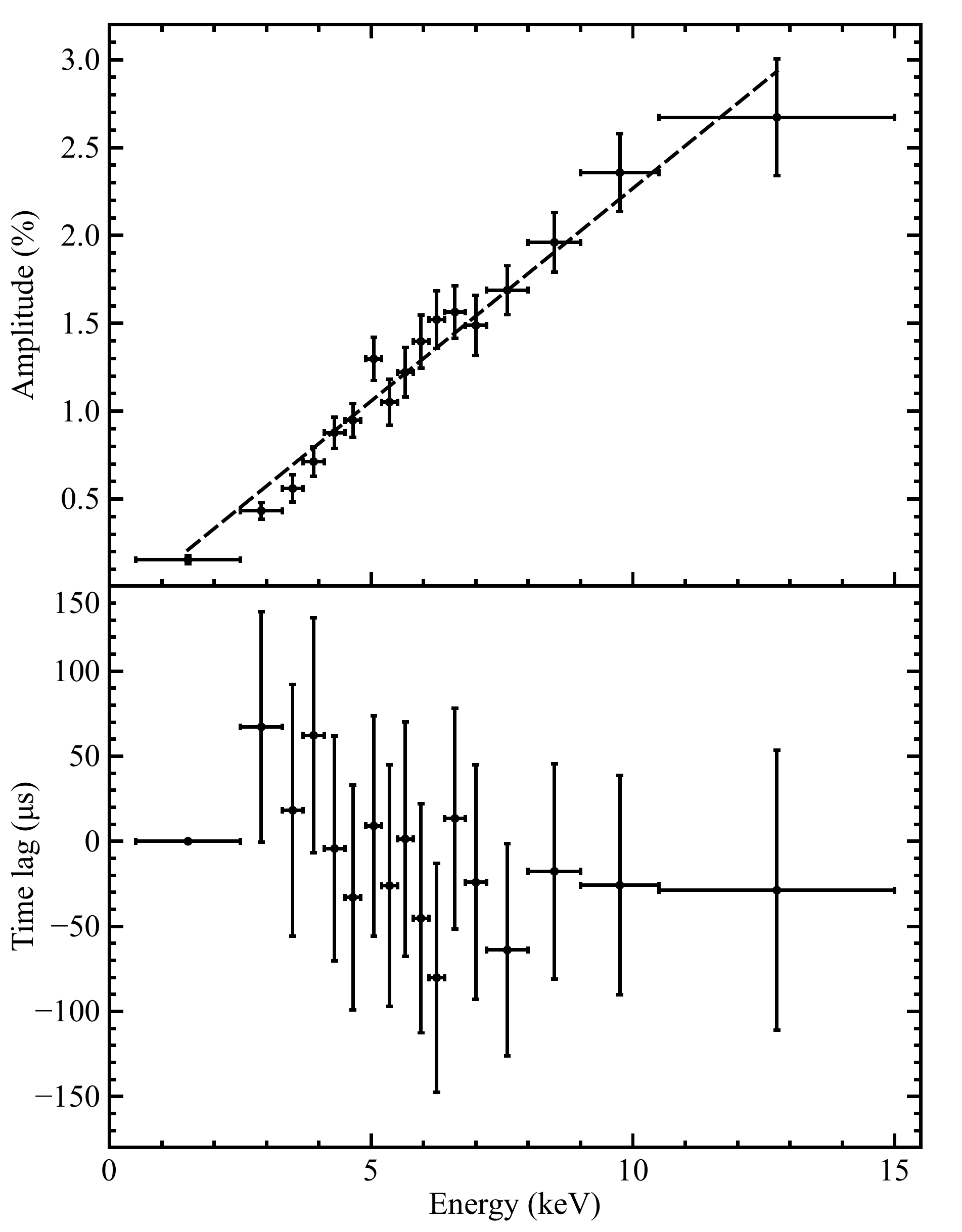}
\caption{\textit{Top panel -} Evolution of the pulse profile fractional amplitude obtained from the PN observation as a function of energy, and best-fitting model to the data (dashed line). \textit{Bottom panel -} Time lags in $\mu s$ as a function of energy, calculated with respect to the first energy band.}
\label{fig:amp_vs_energy}
\end{figure}

We modelled the temporal evolution of the pulse phase delays with the relation:

\begin{eqnarray}
\label{eq:ph}
\Delta \phi(t)=\phi_0+\Delta \nu_0\,(t-T_0)+R_{orb}(t),
\end{eqnarray}
where $T_0$ represents the reference epoch for the timing solution, $\Delta \nu_0=(\nu_0-\bar{\nu})$ is the difference between the frequency at the reference epoch and the spin frequency used to epoch-fold the data, and $R_{orb}$ is the phase residual caused by differences between the \emph{correct} set of orbital parameters and those used to correct the photon time of arrivals \citep[see e.g.][]{Deeter81}. If a new set of orbital parameters is found, photon time of arrivals are corrected using Eq.~\ref{eq:bary} and pulse phase delays are created and modelled with Eq.~\ref{eq:ph}. This process is repeated until no significant differential corrections are found for the parameters of the model. Obtained best-fit parameters are shown in Tab.~\ref{tab:solution}, while in Fig.~\ref{fig:phase_fit} we report the pulse phase delays with the best-fitting model (top panel), and the residuals with respect to the model. The value of $\tilde{\chi}^2\sim1$ (with 11 degrees of freedom) combined with the distribution of the residuals around zero, clearly show a good agreement between the model and the pulse phase delays.
We investigated the dependence of the pulse profile as a function of energy, dividing the energy range between 0.5 keV to 15 keV into 17 intervals and measuring the fractional amplitude and the time lags of the pulse profile. We adjusted the width of the energy bins considered for the analysis in order to be able to significantly detect the pulsation. Fig.~\ref{fig:amp_vs_energy} shows the dependence of the fractional amplitude of the pulse profile (top panel), and the time lags with respect to the first energy band (bottom panel), both as a function of energy. The fractional amplitude increases from $\sim0.1\%$ at around 2 keV up to $\sim3\%$ at 13 keV. A linear correlation between pulse amplitude and energy is quite clear from the plot; we emphasised that by plotting the best-fitting linear function on top of the data (see dashed line on Fig.~\ref{fig:amp_vs_energy}), corresponding to a slope of $(0.24\pm0.1)\%$ keV$^{-1}$. On the other hand, no significant time lags are measured, with all the measurements being consistent with a zero lag with respect to the chosen reference profile.

\subsection{Timing analysis of the previous outbursts}

\begin{figure*}
\centering
\includegraphics[width=1\textwidth]{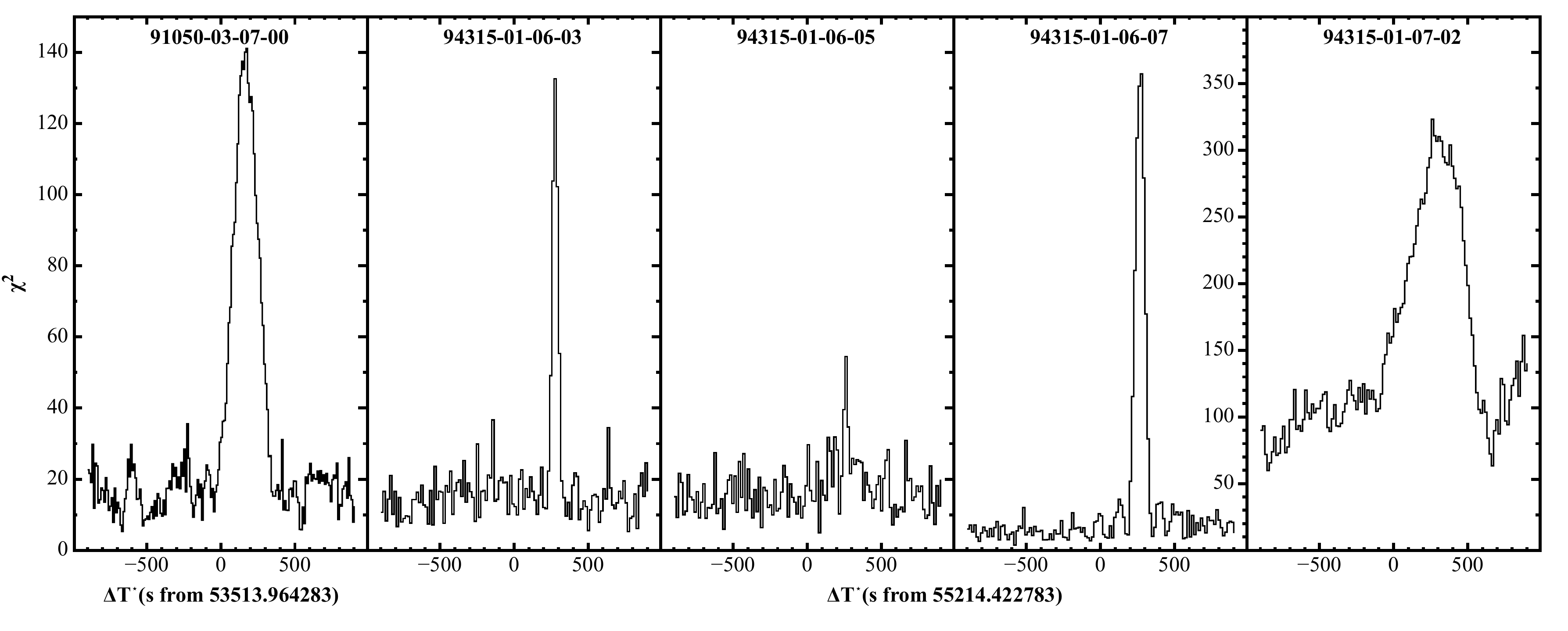}
\caption{Maximum value of $\chi^2$ from the epoch-folding search on the single \textit{RXTE} observations as a function of the $T^{\star}$ values used to correct for the orbital modulation. $\Delta T^{\star}$ represents the delay in seconds from the predicted $T^{\star}$ extrapolated from the timing solution obtained analysing the 2001 outburst. In order to identify the different outbursts in the panels we reported the reference time MJD 53513.964283 and MJD 55214.422783 for the 2005 and 2010 outburst, respectively.}
\label{fig:tstars}
\end{figure*}

\begin{figure}
\centering
\includegraphics[width=0.48\textwidth]{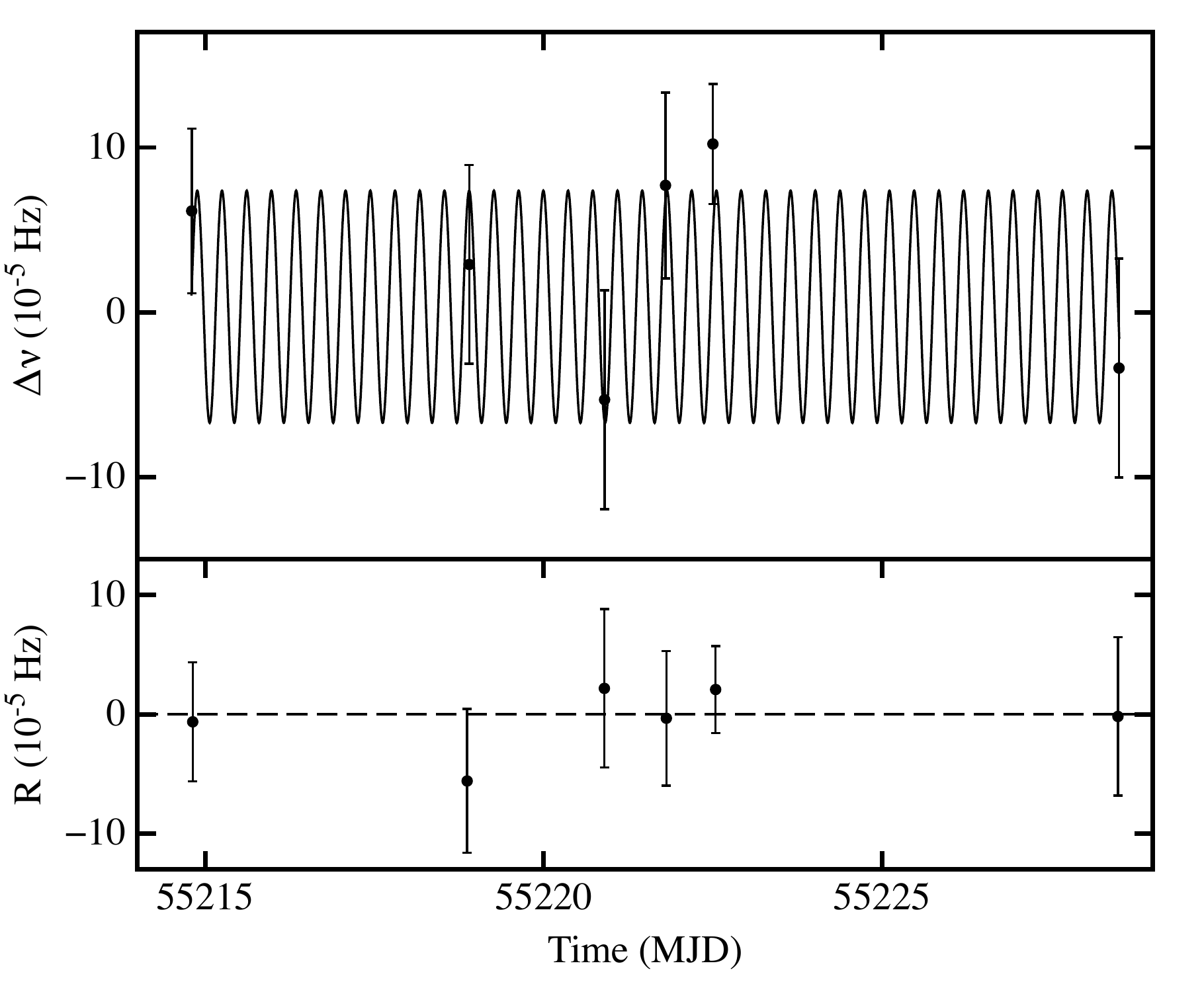}
\caption{\textit{Top panel -} Pulse frequency evolution during the 2010 outburst of the source. The solid line represents the best-fitting orbital model. \textit{Bottom panel -} Residuals in units of $10^{-5}$ Hz with respect to the best-fitting model. }
\label{fig:doppler}
\end{figure}

Using the previously described $T^\star$ searching technique we investigated the whole available \textit{RXTE} dataset of SAX J1748.9$-$2021 to search for more pulsation episodes (see Tab.~\ref{tab:obs}), with the exception of the 2001 outburst (Obs ID P960035 and P960084) for which has been already reported an accurate timing solution (P09). Starting from the set of orbital parameters reported by P09 and using Eq.~\ref{eq:barygen}, we corrected photon times of arrival varying $T^\star$ in order to explore all possible values for the parameter. For each set of orbital parameters we epoch-folded the data sampling the signal with 16 phase bins. Given the correlation between $T^\star$ and spin frequency on time scales shorter than the orbital period and given the low achievable accuracy for the spin frequency in the single \textit{RXTE} observations due to their relative short lengths, we decided to epoch-fold the data fixing the spin frequency value to the one reported by P09. We found evidence for pulsations in five observations, one corresponding to the 2005 outburst \citep[already reported by][]{Gavriil07,Altamirano08a, Patruno09a}, and four corresponding to the 2010 outburst \citep[some of them already reported by][]{Patruno10a}. Fig.~\ref{fig:tstars} shows, for each observation, the $\chi^2$ curve as a function of the $T^\star$ adopted to correct the photon times of arrival. The detected pulsations correspond to $T^\star$ consistent, within the error estimated by applying Eq.~\ref{eq:sigmaTstar}, with respect to the predicted values extrapolated from the solution reported by P09. \\
From Fig.~\ref{fig:tstars} we note that the $\chi^2$ distribution as a function of $\Delta T^\star$ showed a broad range of the full width at half-maximum values, going from $\sim40$ seconds up to $\sim300$ seconds. This can be explained taking into account that the Doppler shift effect (caused by the binary orbital motion), varies in intensity as a function of the orbital phase of the source. Hence, if investigated on time scales shorter than the orbital period, the signal-to-noise ratio of the pulsation can be more or less sensitive to variations of the orbital parameters depending on the orbital position. If, for instance, the source is at an orbital phase where the Doppler effects are relatively intense, small changes of the orbital parameters will strongly degrade the signal (i.e., Fig.~\ref{fig:tstars} Obs ID 94315-01-06-07). On the other hand, if the Doppler effects are relatively weak, large variations of the orbital parameters are required to degrade the signal (i.e., Fig.~\ref{fig:tstars} Obs ID 94315-01-07-02).
In Tab.~\ref{tab:solution} we reported the value of $T^\star$ and the source spin frequency derived from the pulse detection of the 2005 outburst. The $1\sigma$ uncertainties associated to the parameters have been derived by means of Monte Carlo simulations.\\
In spite of the four pulsation episodes observed during the 2010 outburst, the time gap between the detections as well as their statistics did not allow us to perform a phase-coherent timing analysis. However, we managed to study the pulse frequency drift using epoch-folding search techniques with the aim of investigating the orbital parameters of the source. We proceeded as follows:  1) using the four observations reported in Fig.~\ref{fig:tstars} we estimated an averaged value for $T^\star=55214.4259$ MJD, 2) we corrected photon times of arrival using the solution reported by P09, except for the $T^\star$ parameter for which we used the aforementioned value, 3) we performed an epoch-folding search around the spin frequency of the source for each of the available \textit{RXTE} observations of the 2010 outburst. As a result we detected the pulsation in two additional observations, corresponding to Obs ID 94315-01-05-01 (MJD 55214.815) and Obs ID 94315-01-06-01 (MJD 55218.869), both with an exposure of $\sim1.6$ ks. To model the behaviour of the the spin frequency with time we used the expression $\nu = \nu_0(1-\dot{z}(t)/c)$. In order to investigate the differential corrections on the orbital parameters we differentiated the former relation finding the expression:
\begin{eqnarray}
\begin{split}
\nu(t) =& (\nu_0 + \delta\nu_0) \big\{1 - \frac{2\pi}{P_{orb}} \big[\cos(l(t)) \delta x + \frac{x}{P_{orb}} \big(\cos(l(t)) \\
 - & l(t)  \sin(l(t))\big) \delta P_{orb} - \frac{2\pi}{P_{orb}} x  \sin(l(t))  \delta T^\star \big] \big\},
\end{split}
\end{eqnarray}
where $\nu_0$ is the average spin frequency, $\delta\nu_0$ is the differential correction to the spin frequency, $l(t)=2\pi/P_{orb}(t-T^\star)$ is the mean longitude, $\delta x$, $\delta P_{orb}$ and $\delta T^\star$ represent the differential corrections to the orbital parameters. Given the limited number of points we restricted our search to $\delta T^\star$ and  $\delta\nu_0$. Fig.~\ref{fig:doppler} shows the pulse frequency as a function of time for the 2010 outburst as well as the best-fit orbital model. Best-fit parameters are reported in Tab.~\ref{tab:solution}.

\begin{table*}

\begin{tabular}{l | c  c  c  c}
Parameters             & 2001 outburst & 2005 outburst & 2010 outburst & 2015 outburst \\
\hline
\hline
R.A. (J2000) &  \multicolumn{4}{c}{$17^h48^m52^s.163$}\\
DEC (J200) & \multicolumn{4}{c}{$-20^\circ21'32''.40$}\\
Orbital period $P_{orb}$ (s) & 31554.9(1) & $-$$^a$ & $-$$^a$ & 31555.3(3)\\
Projected semi-major axis a sin\textit{i/c} (lt-ms) &387.60(4) & $-$$^a$& $-$$^a$ & 387.57(2) \\
Ascending node passage $T^{\star}$ (MJD) & 52191.507190(4) & 53513.9661(1)& 55214.42571(3)& 57085.444718(9)\\
Eccentricity (e) &$ < 2.3 \times 10^{-4} $&  $-$$^a$& $-$$^a$&$< 8\times 10^{-5}$\\
Spin frequency $\nu_0$ (Hz) &442.36108118(5)& 442.36108(4)& 442.36111(2)& 442.3610957(2)\\
Epoch of $\nu_0$ and $\dot{\nu}_0$, $T_0$ (MJD) & 52190.0 & 53535.4& 55214.7& 57085.7 \\
\hline
$\chi^2_\nu$/d.o.f. & $-$ & $-$& 4.6/3& 10.9/11\\
\end{tabular}
\caption{Orbital parameters of SAX J1748.9$-$2021 obtained by analysing the 2001 outburst \citep[first column;][]{Patruno09a}, 2005, 2010 and  2015 outbursts investigated in this work (second, third, and fourth column). Errors are at 1$\sigma$ confidence level. The reported X-ray position of the source has a pointing uncertainty of 0.6$''$ \citep[see e.g.][]{in-t-Zand01, Pooley02}. $^a$ This parameter has been fixed to the value obtained from the 2001 outburst timing solution.}
\label{tab:solution}
\end{table*}

\section{Discussion}

We have presented an updated timing solution for the intermittent accreting millisecond X-ray pulsar SAX J1748.9$-$2021 obtained by phase connecting the pulsations detected during the \textit{XMM-Newton} observation of its 2015 outburst. The new set of orbital parameters is compatible within the errors with the previous timing solution obtained from the analysis of the 2001 outburst (P09).

\subsection{The spin evolution of SAX J1748.9$-$2021}
As already discussed in previous sections, the source has been observed in outburst 5 times since its discovery, and in 4 of them X-ray pulsation has been detected. Tab.~\ref{tab:solution} shows that we have accurate measurements of the spin frequency only for the 2001 and 2015 outbursts. The difference in frequency between these outbursts is $\Delta \nu = \nu_{2015}-\nu_{2001} = (1.45\pm0.02)\times 10^{-5}$ Hz, where the error quoted is the statistical error obtained propagating those reported in Tab.~\ref{tab:solution}. The variation of spin frequency between the outbursts suggests a significant spin-up of the NS. This trend cannot be confirmed by the rest of the outbursts, because the uncertainties on the spin frequency values estimated from the 2005 and 2010 outbursts are large enough to be consistent both with $\nu_{2001}$ and $\nu_{2015}$. We can speculatively discuss the possibility that the $\Delta \nu$ observed is the result of the accretion torque exerted on the NS as a consequence of the mass transfer from the companion. A rough estimation of the spin-up frequency derivative, $\dot{\nu}$, can be obtained by dividing the spin difference $\Delta \nu$ by the time interval where the source appeared in outburst. Combining Fig.~1 of \citet{Altamirano08a}, the light-curve of the 2010 outburst (not shown in this work) and Fig.~\ref{fig:lc} in this paper, we can estimate that between the first detected X-ray pulsations (2001) and the latest one (2015), the source spent almost 170 days in outburst, corresponding to $\dot{\nu}\simeq 1\times 10^{-12}$ Hz/s. It is interesting to note that, although very approximate, the frequency derivative obtained is in line with values measured in other AMXPs such as IGR J00291$+$5934 \citep{Burderi10}, SAX J1808.4$-$3658 \citep{Burderi06}, XTE J1807$-$294 \citep{Riggio08}, and IGR J17480$-$2446 \citep{Papitto11a}. Finally, with some assumptions on the accretion torque modelling \citep[see][for a detailed dissertation of the subject]{Burderi07} we can estimate the NS magnetic field as:

\begin{eqnarray}
B_9\simeq 1.6\,\phi^{-7/4}_{0.5}I^{7/2}_{45}R^{-6}_6m^{3/2}\dot{\nu}^{7/2}_{12}L^{-3}_{37}
\end{eqnarray}
where $B_9$ is the NS magnetic field in units of $10^9$ Gauss, $\phi_{0.5}$ is a model-dependent dimensionless number usually between 0 and 1 \citep{Ghosh79b,Wang96,Burderi98b} in units of 0.5, $I_{45}$ is the moment of inertia of the NS in units of $10^{45}$ g cm$^2$, $R_6$ is the NS radius in units of $10^6$ cm, $m=1.4$ is the NS mass in solar masses, $\dot{\nu}_{12}$ is the spin frequency derivative in units of $10^{-12}$ Hz/s and $L_{37}$ is the luminosity of the source in units of $5\times 10^{37}$ erg s$^{-1}$ corresponding to the 1$-$50 keV unabsorbed source luminosity measured combining \textit{XMM-Newton} and \textit{INTEGRAL} observations (assuming a distance of 8.5 kpc) during the latest outburst \citep{Pintore2016a}. This value is in agreement with the estimation $B\gtrsim1.3\times 10^{8}$ G reported by \citet{Altamirano08b} from the 2001 outburst of the source.

\subsection{Pulse energy dependence}

An interesting aspect of SAX J1748.9$-$2021 is the behaviour of its pulse profile as function of energy. As shown in Fig.~\ref{fig:amp_vs_energy}, the pulse fractional amplitude clearly increases with energy, varying from 0.1\% up to 2.5\% in the energy range 0.5$-$15 keV, confirming the findings of P09 obtained with \textit{RXTE} during the 2001 outburst of the source. A similar behaviour has been observed in several AMXPs such as Aql X-1 \citep{Casella08}, SWIFT J1756.9$-$2508 \citep{Patruno10b}, XTE J1807$-$294 \citep{Kirsch04}. An increase of the fractional amplitude with energy has been detected also in IGR J00291+5934 \citep{Falanga05b}, although here the energy dependence is more complex. The origin of the phenomenon is still unclear, however mechanisms such as strong Comptonisation of the beamed radiation have been proposed to explain the hard spectrum of the pulsation observed in these sources \citep{Falanga07b}. An alternative scenario proposed by \citet{Muno02, Muno03}, and reported by P09 to describe the behaviour of the SAX J1748.9$-$2021, attempts to explain the X-ray pulsations with the presence of a hot spot region emitting as a blackbody with a temperature significantly different with respect to the NS surface. Such a configuration could explain the increasing pulse amplitude with energy in the observer rest frame. However, it is interesting to note that other AMXPs such as SAX J1808.4$-$3658 \citep{Cui98b, Falanga07b}, XTE J1751-305 \citep{Falanga07b} and IGR J17511$-$3057 \citep{Falanga11} show the exact opposite correlation between pulse fractional amplitude and energy. 

\subsection{Orbital period evolution} 
\begin{figure}
\centering
\includegraphics[width=0.5\textwidth]{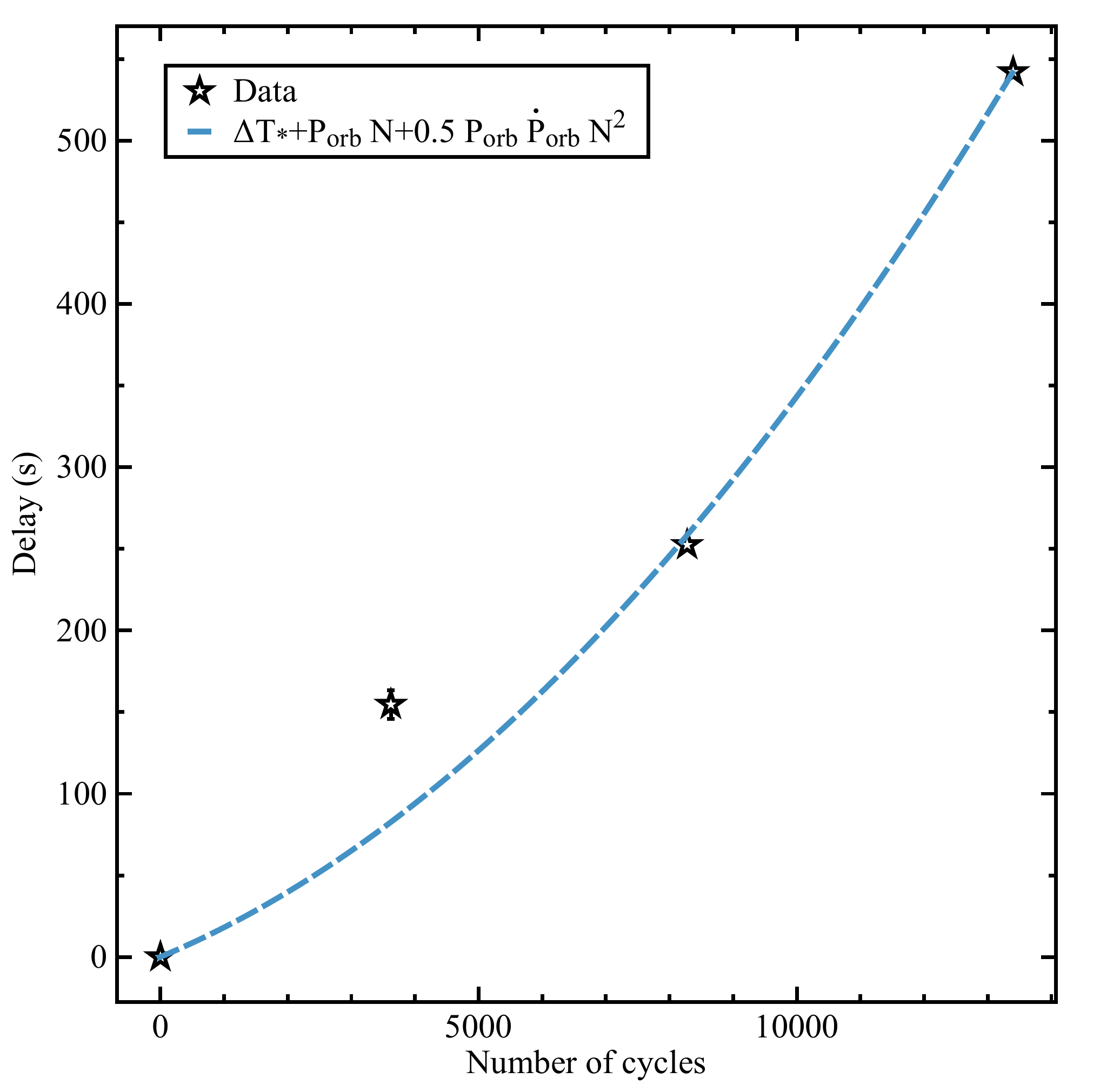}
\caption{Differential correction on the NS time of passage from the ascending node for each of the outbursts showing X-ray pulsations. The cyan dashed line represents the best-fitting parabola used to model the data. }
\label{fig:fit_tstar2}
\end{figure}

As reported in Tab.~\ref{tab:solution} we have measurements of the time of passage of the NS at the ascending node for 4 out of the 5 observed outbursts of the source.  We note that the correction to the predicted $T^\star_{predict}=T^\star_{2001}+N P_{orb_{2001}}$, increases with time. Here, the integer $N$ represents the number of orbital cycles elapsed between two different $T^\star$ \citep[see e.g.,][]{diSalvo08, Burderi09}. In Fig.~\ref{fig:fit_tstar2} we report the differential correction on the NS passage from the ascending node (with respect to the timing solution of P09) for each of the outbursts where we detected the pulsation, as a function of the orbital cycles elapsed from the reference time. We fitted the data with the expression: 
\begin{eqnarray}
\label{eq:fit_tstar}
\Delta T^\star = \delta T^\star_{2001} + N\, \delta P_{orb_{2001}}+0.5\,N^2\, \dot{P}_{orb}P_{orb_{2001}},
\end{eqnarray} 
where the correction to the adopted time of passage from the ascending node, $\delta T^\star_{2001}$, the correction to the orbital period, $\delta P_{orb_{2001}}$, and the orbital-period derivative, $\dot{P}_{orb}$, are the fit parameters. 
We found the best-fitting values $\delta T^\star_{2001}=(0.05\pm0.35)$ MJD, $\delta P_{orb_{2001}}=(0.0163\pm0.0008)$ s, and $\dot{P}_{orb}=(1.14\pm0.04)\times 10^{-10}$ s/s, with a $\chi^2=78.4$ (for 1 d.o.f.). We note that the large $\chi^2$ value is influenced by the $T^\star$ value estimated from the 2005 outburst, that differs more than $10\sigma$ from the best-fitting model. We remind the reader that the aforementioned parameter has been deduced from a single short observation ($\sim 1$ ks of data) during the whole outburst. The best fit of the $T^\star$ evolution is clearly statistically not acceptable, likely reflecting a complex orbital period evolution that we can not investigate with such data, or an underestimation of the statistical uncertainties. However, under the simple hypothesis that the underlying evolution of $T^\star$ with time is compatible with Eq.~\ref{eq:fit_tstar}, we can re-modulate the uncertainty on the fitting parameters taking into account the root-mean-square of the fit residuals, i.e. multiplying the fit uncertainties by the square root of the $\chi^2$ per degree of freedom, that in this specific case corresponds to a factor $\sim9$. We can re-write the best-fitting parameters as $\delta T^\star_{2001}=(0.05\pm3.1)$ MJD, $\delta P_{orb_{2001}}=(0.0163\pm0.007)$ s, and $\dot{P}_{orb}=(1.1\pm0.3)\times 10^{-10}$ s/s. We find no significant correction for $T^\star_{2001}$ and a marginally significant correction for $P_{orb_{2001}}$. On the other hand we find, for the first time for this source, a marginally significant ($3.5\sigma$) orbital-period derivative, which suggests a very rapid increase of the orbital period. 

Under the hypothesis that the value of $\dot{P}_{orb}$ is reliable, we can investigate the orbital evolution of the system. As a first step we can estimate the mass-loss rate from the secondary expected from the observed orbital period and the orbital-period derivative. Following \citet{Burderi10}, we can write the averaged secondary mass-loss rate as:
\begin{eqnarray}
\label{eq:m2_pdot}
\dot{m}_{est,-8}=1.9 \times (3n-1)^{-1}\,m_{2,0.1}\Big(\frac{\dot{P}_{orb,-10}}{P_{orb,9h}}\Big),
\end{eqnarray}
where $\dot{m}_{est,-8}$ is expressed in units of $10^{-8}$ M$_{\odot}/yr$, $n$ is the index of the mass-radius relation of the secondary R$_2\propto$ M$_2^n$, $m_{2,0.1}$ is the mass of the companion star in units of 0.1 M$_{\odot}$, $\dot{P}_{orb,-10}$ is the orbital-period derivative in units of $10^{-10}$ s/s, and $P_{orb,9h}$ is the orbital period in units of 9 hours (appropriate for SAX J1748.9$-$2021 since $P_{orb}=8.76$ h). We remind the reader that the previous relation is valid for small mass ratios, $q=m_2/m_1\leq 0.8$, since it assumes the Paczy\'nski approximation \citep{Paczynski71} to describe the secondary Roche Lobe.

Even though the mass of the companion star is still unknown, we can at least define a mass range by means of the binary mass function 
\begin{eqnarray}
f(m_1,m_2)=\frac{4\pi^2(a \sin i)^3}{G P^2_{orb}}=\frac{(m_2 \sin i)^3}{(m_1+m_2)^2},
\end{eqnarray}

where $G$ is the gravitational constant, and $i$ is the inclination angle of the binary. Adopting the orbital parameters reported in Tab.~\ref{tab:solution}, assuming $i\leq60^\circ$ (taking into account that no X-ray eclipses, neither dips have been observed), and a NS mass $m_1=1.4$ M$_{\odot}$ we estimate $m_2 \geq 0.12$ M$_{\odot}$. Moreover, an upper limit for the companion star can be estimated under the assumption that a Roche lobe-filling star is close to the lower main sequence, that translates in the relation $m_2\simeq 0.11 P_{orb,h}$ M$_{\odot}$ \citep{King88}, and corresponds to $\simeq1$ M$_{\odot}$ for this source. We note that the latter value corresponds to the mass value estimated by \citet{Altamirano08a}. 

\begin{figure}
\centering
\includegraphics[width=0.5\textwidth]{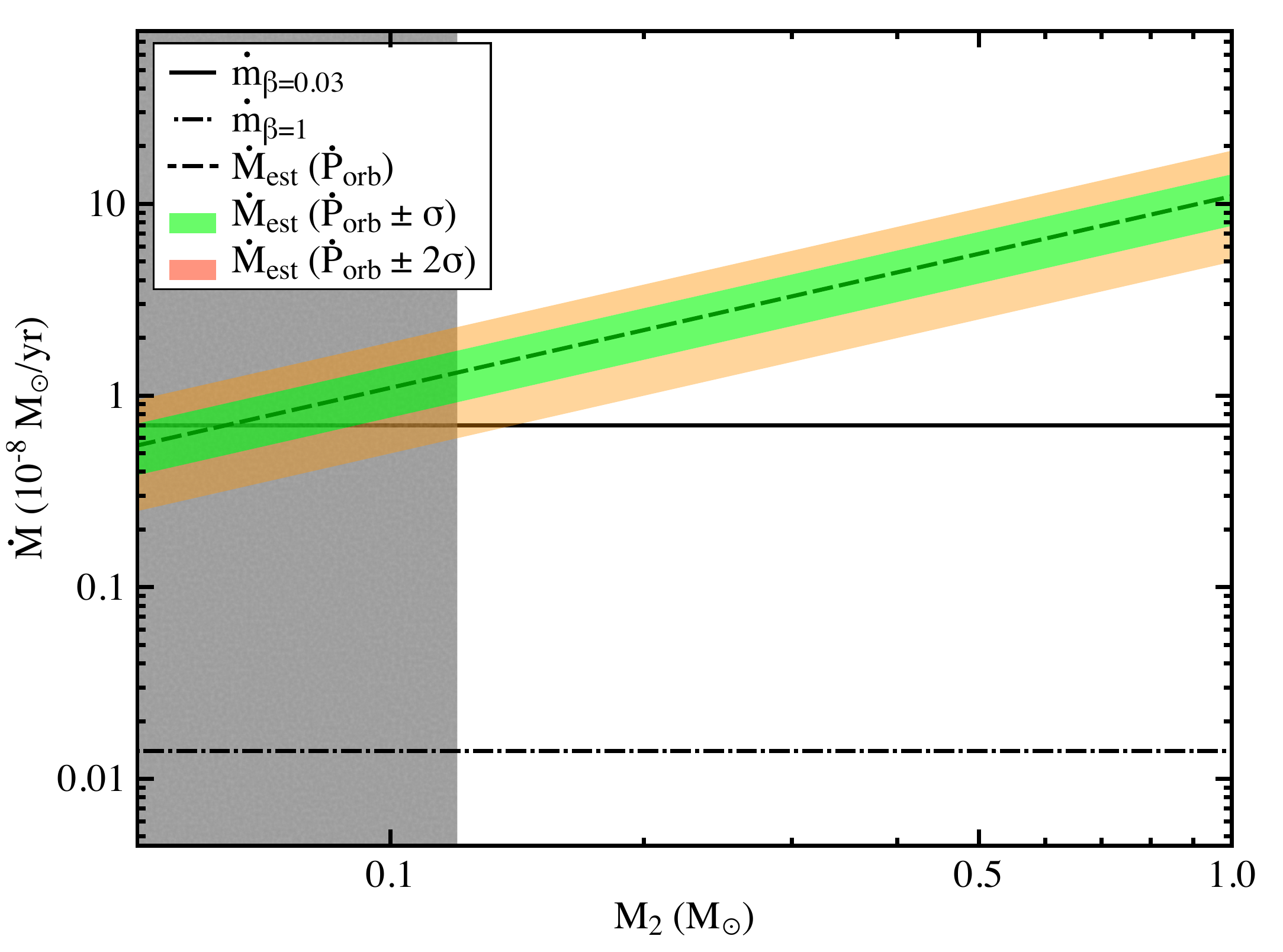}
\caption{Mass-loss rate of the companion of SAX J1748.9$-$2021 (dashed line) estimated from the measured orbital period derivative as a function of its mass value. The light-green and light orange regions represent the mass-loss rate estimate taking into account the $1\sigma$ and $2\sigma$ uncertainty of $\dot{P}_{orb}$, respectively. The solid line shows the maximum mass-accretion rate observed during all outbursts of SAX J1748.9$-$2021 (measured during the peak of the 2005 outburst), while the dot-dashed line shows the averaged mass-accretion rate estimated for the conservative mass transfer scenario ($\beta=1$). Finally the dark shaded area represents the constraints on the companion mass imposed by the binary mass function of the system.}
\label{fig:orb_ev}
\end{figure}

Using Eq.~\ref{eq:m2_pdot} we estimate the expected secondary mass-loss rate for mass values of the secondary ranging between 0.12$-$1 M$_\odot$. Regarding the value of the mass-radius index, we note that the observed $\dot{P}_{orb}>0$ (under the assumption that the measured orbital-period derivative reflects the secular evolution of the system) likely implies a companion star in non-thermal equilibrium, with mass lower than 0.3 M$_{\odot}$, and mass-radius relation inverted \citep{King88, Verbunt93}. Assuming a fully convective companion star, and an orbital evolution driver by GR, we substitute $n=-1/3$ in Eq.~\ref{eq:m2_pdot}. As shown in Fig.~\ref{fig:orb_ev} (dashed line), the mass-loss rate varies from $\sim 1.3\times 10^{-8}$ M$_\odot$/yr ($m_2=0.12$ M$_\odot$) up to $\sim 11\times 10^{-8}$ M$_\odot$/yr ($m_2=1$ M$_\odot$). Starting from these numbers we can explore two possible evolutionary scenarios invoking \emph{conservative} and \emph{non-conservative} mass transfer between the secondary and the NS. 

The first scenario assumes that mass transferred from the companion star during the outburst must be completely accreted by the NS, while during the quiescence states no mass is accreted or lost from the system. Defining $\beta$ as the fraction of the mass transferred from the secondary to the NS, we can identify the conservative scenario with $\beta=1$. To verify whether the inferred mass-loss rate reported in Fig.~\ref{fig:orb_ev} is somehow compatible with the conservative scenario we need to compare it with the averaged mass-transfer rate extrapolated from the averaged observed flux of the source. Assuming the unabsorbed bolometric luminosity observed from the \textit{XMM-Newton} observation \citep[$\sim 5.7\times 10^{-37}$ erg/s, see][]{Pintore2016a} as the averaged luminosity during the outburst, and taking into account that SAX J1748.9$-$2021 spends roughly 60 days in outburst every 5 years, we can infer the averaged mass-loss rate $\dot{m}_{\beta=1}\sim1.4\times 10^{-10}$ M$_\odot$/yr. As clearly shown in Fig.~\ref{fig:orb_ev}, $\dot{m}_{\beta=1}$ (dot-dashed line) and the expected secondary mass-loss rate estimated from the orbital-period derivative are not compatible for any reasonable explored value of the mass of the companion star. This result strongly suggests that the orbital evolution of the systems (characterised by the observed $\dot{P}_{orb}$), cannot be described by a conservative mass-transfer scenario.

The second scenario assumes a non-conservative mass transfer ($\beta < 1$), meaning that also during the quiescence phases the companion star fills its Roche lobe but instead of being accreted onto the NS the matter is ejected from the system. A rough estimate of $\beta$ for SAX J1748.9$-$2021 can be obtained by computing the outburst phase duty cycle, corresponding to roughly 60 days every 5 years, hence $\beta\sim 3\%$. To investigate this scenario we make the assumption that the NS accretes at a rate equal to the one shown during the peak of the most luminous outburst of the source \citep[2005 outburst, see][]{Altamirano08b}, that corresponds to $\dot{m}_{\beta=0.03}\sim 7\times 10^{-9}$ M$_{\odot}$/yr (the value has been extrapolated from the unabsorbed bolometric flux measured from the \textit{RXTE} observation of the source). It is interesting to note from Fig.~\ref{fig:orb_ev}, that $\dot{m}_{\beta=0.03}$ (solid line) and the estimated secondary mass-loss rate are compatible for values of the companion star in the range 0.05$-$0.09 M$_{\odot}$ ($1\sigma$ confidence level) and 0.04$-$0.14 M$_{\odot}$ ($2\sigma$ confidence level). We find that the $2\sigma$ interval is compatible with the lower limit of $m_2$ estimated from the binary mass function. Our finding, taking into account all the assumptions and the caveats explained above, suggests that large value of the orbital-period derivative reported for SAX J1748.9$-$2021 reflects an highly non-conservative mass transfer scenario for which a large amount of mass lost from the companion star is expelled from the system. 

The last step needed to investigate the orbital evolution of the source is to verify whether the non-conservative mass-transfer unveiled in the previous paragraph is compatible with a secular evolution. Following \citet{diSalvo08}, we solve the binary evolutionary equation under the assumptions that: 1) the angular momentum losses are driven by emission of gravitational waves; 2) the secondary mass-radius relation is $R_2\propto M^{n}_2$; and 3) the NS accretes mass through Roche lobe overflow, meaning that the evolution of the secondary mass radius $\dot{R}_2/R_2$ can be expressed in terms of to the evolution of the of the secondary Roche lobe $\dot{R}_{L2}/R_{L2}$, where for $R_{L2}$ we adopt the approximation $R_{L2}=2/3^{4/3}[q/(1+q)]^{1/3}a$ \citep{Paczynski71}. Consequently, we can write the orbital-period derivative due to general relativity (GR) as:
\begin{eqnarray}
\dot{P}_{orb,-13}=-1.1\times\Big[\frac{n-1/3}{n-1/3+2g}\Big]m_1\,m_2\,m^{-1/3}\,P^{-5/3}_{orb,9h}
\end{eqnarray}    
where $\dot{P}_{orb,-13}$ is expressed in units of $10^{-13}$ s/s, $g=1-\beta q-(1-\beta)(\alpha+q/3)/(1+q)$ reflects the angular momentum losses because of mass loss from the system, and $\alpha=l_{ej}P_{orb}\,m^2/(2\pi a^2 m_1^2)$ is the specific angular momentum of the matter leaving the system ($l_{ej}$) in units of the specific angular momentum of the companion star located at a distance $r_2$ from the center of mass of the system and with an orbital separation $a$. 
Adopting $n=-1/3$, $m_1=1.4$ M$_\odot$, $m_2=0.12$ M$_\odot$ (compatible within $2\sigma$ with the mass value with our finding described above), we find that, in order to obtain an orbital-period derivative consistent with values determined from the timing analysis ($\dot{P}_{orb}=1.1\times 10^{-10}$s/s), the specific angular momentum of the matter leaving the system must be $\alpha\simeq0.7$. We note that this value is close to the specific angular momentum of the matter at the inner Lagrangian point $\alpha=[1-0.462(1+q)^{2/3}q^{1/3}]^2\sim0.63$. Therefore, we can conclude suggesting that the large orbital-period derivative observed in SAX J1748.9$-$2021 is compatible with an highly non-conservative GR driven mass transfer, with matter leaving the system in the proximity of the inner Lagrangian point. \cite{diSalvo08} and \cite{Burderi09} reported the same phenomenon for the AMXP SAX J1808.4$-$3658, proposing the \emph{radio-ejection} mechanism as possible explanation.

We note that for the case of SAX J1808.4$-$3658 alternative models have been proposed to explain the orbital period phenomenology. \citet{Hartman08, Hartman09b} and \citet{Patruno12a} suggested that the observed orbital period derivative might instead reflect short-term interchange of angular momentum between the companion star and the orbit. In this scenario the variable gravitational quadrupole moment (GQC) of the companion star (which generates by cyclic spin-up and spin-down on its outer layers) should be responsible for the orbital period changes \citep{Applegate1992a, Applegate1994a}. This mechanism has been applied to describe the time evolution of the orbital period observed in the black widow systems PSR B1957$+$20 \citep{Arzoumanian1994a, Applegate1994a}, PSR J2051$-$0827 \citep{Doroshenko2001a, Lazaridis2011a}, in the redback system PSR J2339$-$0533 \citep{Pletsch2015a}, and in the transitional redback system PSR J1023$+$0038 \citep{Archibald2013a}. 
\\
In the case of SAX J1748.9$-$2021, the present data do not allow us to constrain a second time derivative or a sinusoidal trend of the orbital period, and therefore it is not clear if the orbital period derivative will cyclicly change with time. We cannot exclude, however, that the orbital period of SAX J1748.9$-$2021 might exhibit more complex behaviour on longer time scales. Nonetheless, in all the above mentioned systems where GQC has been invoked, the orbital period derivatives vary on timescales $\lesssim 10$ yr, almost a factor of 2 shorter than time interval studied in this work. Furthermore, we note that this mechanism increases in efficiency for Roche-lobe filling factors of the companion star lower than unity \citep[as a consequence of the strong dependence on the ratio $R_{L2}/R_2$;][]{Applegate1994a}.  
\\
In conclusion, we propose that the large orbital period derivative observed for SAX J1748.9$-$2021 reflects a highly non-conservative mass transfer where almost 97\% of the matter is ejected from the system with the specific angular momentum of the inner Lagrangian point. However, given the level of significance for the reported orbital-period derivative, more X-ray outbursts are required to further investigate the secular orbit evolution of SAX J1748.9$-$2021 in order to confirm or disprove the proposed scenario.

\section*{Acknowledgments}
We thank N. Schartel for the possibility to perform the ToO observation in the Director Discretionary Time, and the XMM-Newton team for the support for the observation. We gratefully acknowledge the Sardinia Regional Government for the financial support (P. O. R. Sardegna F.S.E. Operational Programme of the Autonomous Region of Sardinia, European Social Fund 2007-2013 - Axis IV Human Resources, Objective l.3, Line of Activity l.3.1). This work was partially supported by the Regione Autonoma della Sardegna through POR-FSE Sardegna 2007- 2013, L.R. 7/2007, Progetti di Ricerca di Base e Orientata, Project N. CRP-60529. We also acknowledge financial contribution from the agreement ASI-INAF I/037/12/0.The High-Energy Astrophysics Group of Palermo acknowledges support from the Fondo Finalizzato alla Ricerca (FFR) 2012/13, project N. 2012- ATE-0390, founded by the University of Palermo.

\bibliographystyle{mnras}
\bibliography{biblio}

\begin{thebibliography}{}
\makeatletter
\relax
\def\mn@urlcharsother{\let\do\@makeother \do\$\do\&\do\#\do\^\do\_\do\%\do\~}
\def\mn@doi{\begingroup\mn@urlcharsother \@ifnextchar [ {\mn@doi@}
  {\mn@doi@[]}}
\def\mn@doi@[#1]#2{\def\@tempa{#1}\ifx\@tempa\@empty \href
  {http://dx.doi.org/#2} {doi:#2}\else \href {http://dx.doi.org/#2} {#1}\fi
  \endgroup}
\def\mn@eprint#1#2{\mn@eprint@#1:#2::\@nil}
\def\mn@eprint@arXiv#1{\href {http://arxiv.org/abs/#1} {{\tt arXiv:#1}}}
\def\mn@eprint@dblp#1{\href {http://dblp.uni-trier.de/rec/bibtex/#1.xml}
  {dblp:#1}}
\def\mn@eprint@#1:#2:#3:#4\@nil{\def\@tempa {#1}\def\@tempb {#2}\def\@tempc
  {#3}\ifx \@tempc \@empty \let \@tempc \@tempb \let \@tempb \@tempa \fi \ifx
  \@tempb \@empty \def\@tempb {arXiv}\fi \@ifundefined
  {mn@eprint@\@tempb}{\@tempb:\@tempc}{\expandafter \expandafter \csname
  mn@eprint@\@tempb\endcsname \expandafter{\@tempc}}}

\bibitem[\protect\citeauthoryear{{Alpar}, {Cheng}, {Ruderman}  \&
  {Shaham}}{{Alpar} et~al.}{1982}]{Alpar82}
{Alpar} M.~A.,  {Cheng} A.~F.,  {Ruderman} M.~A.,   {Shaham} J.,  1982, \mn@doi
  [\nat] {10.1038/300728a0}, \href
  {http://adsabs.harvard.edu/abs/1982Natur.300..728A} {300, 728}

\bibitem[\protect\citeauthoryear{{Altamirano}, {Casella}, {Patruno}, {Wijnands}
   \& {van der Klis}}{{Altamirano} et~al.}{2008a}]{Altamirano08a}
{Altamirano} D.,  {Casella} P.,  {Patruno} A.,  {Wijnands} R.,   {van der Klis}
  M.,  2008a, \mn@doi [\apjl] {10.1086/528983}, \href
  {http://adsabs.harvard.edu/abs/2008ApJ...674L..45A} {674, L45}

\bibitem[\protect\citeauthoryear{{Altamirano}, {van der Klis}, {M{\'e}ndez},
  {Jonker}, {Klein-Wolt}  \& {Lewin}}{{Altamirano}
  et~al.}{2008b}]{Altamirano08b}
{Altamirano} D.,  {van der Klis} M.,  {M{\'e}ndez} M.,  {Jonker} P.~G.,
  {Klein-Wolt} M.,   {Lewin} W.~H.~G.,  2008b, \mn@doi [\apj] {10.1086/590897},
  \href {http://adsabs.harvard.edu/abs/2008ApJ...685..436A} {685, 436}

\bibitem[\protect\citeauthoryear{{Applegate}}{{Applegate}}{1992}]{Applegate1992a}
{Applegate} J.~H.,  1992, \mn@doi [\apj] {10.1086/170967}, \href
  {http://adsabs.harvard.edu/abs/1992ApJ...385..621A} {385, 621}

\bibitem[\protect\citeauthoryear{{Applegate} \& {Shaham}}{{Applegate} \&
  {Shaham}}{1994}]{Applegate1994a}
{Applegate} J.~H.,  {Shaham} J.,  1994, \mn@doi [\apj] {10.1086/174906}, \href
  {http://adsabs.harvard.edu/abs/1994ApJ...436..312A} {436, 312}

\bibitem[\protect\citeauthoryear{{Archibald}, {Kaspi}, {Hessels}, {Stappers},
  {Janssen}  \& {Lyne}}{{Archibald} et~al.}{2013}]{Archibald2013a}
{Archibald} A.~M.,  {Kaspi} V.~M.,  {Hessels} J.~W.~T.,  {Stappers} B.,
  {Janssen} G.,   {Lyne} A.,  2013, preprint, \href
  {http://adsabs.harvard.edu/abs/2013arXiv1311.5161A} {} (\mn@eprint {arXiv}
  {1311.5161})

\bibitem[\protect\citeauthoryear{{Archibald} et~al.,}{{Archibald}
  et~al.}{2015}]{Archibald2015a}
{Archibald} A.~M.,  et~al., 2015, \mn@doi [\apj] {10.1088/0004-637X/807/1/62},
  \href {http://adsabs.harvard.edu/abs/2015ApJ...807...62A} {807, 62}

\bibitem[\protect\citeauthoryear{{Arzoumanian}, {Fruchter}  \&
  {Taylor}}{{Arzoumanian} et~al.}{1994}]{Arzoumanian1994a}
{Arzoumanian} Z.,  {Fruchter} A.~S.,   {Taylor} J.~H.,  1994, \mn@doi [\apjl]
  {10.1086/187346}, \href {http://adsabs.harvard.edu/abs/1994ApJ...426L..85A}
  {426, 85}

\bibitem[\protect\citeauthoryear{{Bozzo}, {Kuulkers}  \& {Ferrigno}}{{Bozzo}
  et~al.}{2015}]{Bozzo15}
{Bozzo} E.,  {Kuulkers} E.,   {Ferrigno} C.,  2015, The Astronomer's Telegram,
  \href {http://adsabs.harvard.edu/abs/2015ATel.7106....1B} {7106, 1}

\bibitem[\protect\citeauthoryear{{Burderi} \& {Di Salvo}}{{Burderi} \& {Di
  Salvo}}{2013}]{Burderi13}
{Burderi} L.,  {Di Salvo} T.,  2013, \memsai, \href
  {http://adsabs.harvard.edu/abs/2013MmSAI..84..117B} {84, 117}

\bibitem[\protect\citeauthoryear{{Burderi} \& {King}}{{Burderi} \&
  {King}}{1998}]{Burderi98b}
{Burderi} L.,  {King} A.~R.,  1998, \mn@doi [\apjl] {10.1086/311611}, \href
  {http://adsabs.harvard.edu/abs/1998ApJ...505L.135B} {505, L135}

\bibitem[\protect\citeauthoryear{{Burderi}, {Di Salvo}, {Menna}, {Riggio}  \&
  {Papitto}}{{Burderi} et~al.}{2006}]{Burderi06}
{Burderi} L.,  {Di Salvo} T.,  {Menna} M.~T.,  {Riggio} A.,   {Papitto} A.,
  2006, \mn@doi [\apjl] {10.1086/510666}, \href
  {http://adsabs.harvard.edu/abs/2006ApJ...653L.133B} {653, L133}

\bibitem[\protect\citeauthoryear{{Burderi} et~al.,}{{Burderi}
  et~al.}{2007}]{Burderi07}
{Burderi} L.,  et~al., 2007, \mn@doi [\apj] {10.1086/510659}, \href
  {http://adsabs.harvard.edu/abs/2007ApJ...657..961B} {657, 961}

\bibitem[\protect\citeauthoryear{{Burderi}, {Riggio}, {di Salvo}, {Papitto},
  {Menna}, {D'A{\`\i}}  \& {Iaria}}{{Burderi} et~al.}{2009}]{Burderi09}
{Burderi} L.,  {Riggio} A.,  {di Salvo} T.,  {Papitto} A.,  {Menna} M.~T.,
  {D'A{\`\i}} A.,   {Iaria} R.,  2009, \mn@doi [\aap]
  {10.1051/0004-6361/200811542}, \href
  {http://adsabs.harvard.edu/abs/2009A%26A...496L..17B} {496, L17}

\bibitem[\protect\citeauthoryear{{Burderi}, {Di Salvo}, {Riggio}, {Papitto},
  {Iaria}, {D'A{\`\i}}  \& {Menna}}{{Burderi} et~al.}{2010}]{Burderi10}
{Burderi} L.,  {Di Salvo} T.,  {Riggio} A.,  {Papitto} A.,  {Iaria} R.,
  {D'A{\`\i}} A.,   {Menna} M.~T.,  2010, \mn@doi [\aap]
  {10.1051/0004-6361/200912881}, \href
  {http://adsabs.harvard.edu/abs/2010A%26A...515A..44B} {515, A44}

\bibitem[\protect\citeauthoryear{{Casella}, {Altamirano}, {Patruno}, {Wijnands}
   \& {van der Klis}}{{Casella} et~al.}{2008}]{Casella08}
{Casella} P.,  {Altamirano} D.,  {Patruno} A.,  {Wijnands} R.,   {van der Klis}
  M.,  2008, \mn@doi [\apjl] {10.1086/528982}, \href
  {http://adsabs.harvard.edu/abs/2008ApJ...674L..41C} {674, L41}

\bibitem[\protect\citeauthoryear{{Cui}, {Morgan}  \& {Titarchuk}}{{Cui}
  et~al.}{1998}]{Cui98b}
{Cui} W.,  {Morgan} E.~H.,   {Titarchuk} L.~G.,  1998, \mn@doi [\apjl]
  {10.1086/311569}, \href {http://adsabs.harvard.edu/abs/1998ApJ...504L..27C}
  {504, L27}

\bibitem[\protect\citeauthoryear{{Deeter}, {Boynton}  \& {Pravdo}}{{Deeter}
  et~al.}{1981}]{Deeter81}
{Deeter} J.~E.,  {Boynton} P.~E.,   {Pravdo} S.~H.,  1981, \mn@doi [\apj]
  {10.1086/159110}, \href {http://adsabs.harvard.edu/abs/1981ApJ...247.1003D}
  {247, 1003}

\bibitem[\protect\citeauthoryear{{Di Salvo}, {Burderi}, {Riggio}, {Papitto}  \&
  {Menna}}{{Di Salvo} et~al.}{2008}]{diSalvo08}
{Di Salvo} T.,  {Burderi} L.,  {Riggio} A.,  {Papitto} A.,   {Menna} M.~T.,
  2008, \mn@doi [\mnras] {10.1111/j.1365-2966.2008.13709.x}, \href
  {http://adsabs.harvard.edu/abs/2008MNRAS.389.1851D} {389, 1851}

\bibitem[\protect\citeauthoryear{{Doroshenko}, {L{\"o}hmer}, {Kramer},
  {Jessner}, {Wielebinski}, {Lyne}  \& {Lange}}{{Doroshenko}
  et~al.}{2001}]{Doroshenko2001a}
{Doroshenko} O.,  {L{\"o}hmer} O.,  {Kramer} M.,  {Jessner} A.,  {Wielebinski}
  R.,  {Lyne} A.~G.,   {Lange} C.,  2001, \mn@doi [\aap]
  {10.1051/0004-6361:20011349}, \href
  {http://adsabs.harvard.edu/abs/2001A%26A...379..579D} {379, 579}

\bibitem[\protect\citeauthoryear{{Falanga} \& {Titarchuk}}{{Falanga} \&
  {Titarchuk}}{2007}]{Falanga07b}
{Falanga} M.,  {Titarchuk} L.,  2007, \mn@doi [\apj] {10.1086/514805}, \href
  {http://adsabs.harvard.edu/abs/2007ApJ...661.1084F} {661, 1084}

\bibitem[\protect\citeauthoryear{{Falanga} et~al.,}{{Falanga}
  et~al.}{2005}]{Falanga05b}
{Falanga} M.,  et~al., 2005, \mn@doi [\aap] {10.1051/0004-6361:20053472}, \href
  {http://adsabs.harvard.edu/abs/2005A%26A...444...15F} {444, 15}

\bibitem[\protect\citeauthoryear{{Falanga} et~al.,}{{Falanga}
  et~al.}{2011}]{Falanga11}
{Falanga} M.,  et~al., 2011, \mn@doi [\aap] {10.1051/0004-6361/201016240},
  \href {http://adsabs.harvard.edu/abs/2011A%26A...529A..68F} {529, A68}

\bibitem[\protect\citeauthoryear{{Gavriil}, {Strohmayer}, {Swank}  \&
  {Markwardt}}{{Gavriil} et~al.}{2007}]{Gavriil07}
{Gavriil} F.~P.,  {Strohmayer} T.~E.,  {Swank} J.~H.,   {Markwardt} C.~B.,
  2007, \mn@doi [\apjl] {10.1086/523758}, \href
  {http://adsabs.harvard.edu/abs/2007ApJ...669L..29G} {669, L29}

\bibitem[\protect\citeauthoryear{{Ghosh} \& {Lamb}}{{Ghosh} \&
  {Lamb}}{1979}]{Ghosh79b}
{Ghosh} P.,  {Lamb} F.~K.,  1979, \mn@doi [\apj] {10.1086/157498}, \href
  {http://adsabs.harvard.edu/abs/1979ApJ...234..296G} {234, 296}

\bibitem[\protect\citeauthoryear{{Hartman} et~al.,}{{Hartman}
  et~al.}{2008}]{Hartman08}
{Hartman} J.~M.,  et~al., 2008, \mn@doi [\apj] {10.1086/527461}, \href
  {http://adsabs.harvard.edu/abs/2008ApJ...675.1468H} {675, 1468}

\bibitem[\protect\citeauthoryear{{Hartman}, {Patruno}, {Chakrabarty},
  {Markwardt}, {Morgan}, {van der Klis}  \& {Wijnands}}{{Hartman}
  et~al.}{2009}]{Hartman09b}
{Hartman} J.~M.,  {Patruno} A.,  {Chakrabarty} D.,  {Markwardt} C.~B.,
  {Morgan} E.~H.,  {van der Klis} M.,   {Wijnands} R.,  2009, \mn@doi [\apj]
  {10.1088/0004-637X/702/2/1673}, \href
  {http://adsabs.harvard.edu/abs/2009ApJ...702.1673H} {702, 1673}

\bibitem[\protect\citeauthoryear{{Jahoda}, {Markwardt}, {Radeva}, {Rots},
  {Stark}, {Swank}, {Strohmayer}  \& {Zhang}}{{Jahoda} et~al.}{2006}]{Jahoda06}
{Jahoda} K.,  {Markwardt} C.~B.,  {Radeva} Y.,  {Rots} A.~H.,  {Stark} M.~J.,
  {Swank} J.~H.,  {Strohmayer} T.~E.,   {Zhang} W.,  2006, \mn@doi [\apjs]
  {10.1086/500659}, \href {http://adsabs.harvard.edu/abs/2006ApJS..163..401J}
  {163, 401}

\bibitem[\protect\citeauthoryear{{Kaaret}, {Morgan}, {Vanderspek}  \&
  {Tomsick}}{{Kaaret} et~al.}{2006}]{Kaaret06}
{Kaaret} P.,  {Morgan} E.~H.,  {Vanderspek} R.,   {Tomsick} J.~A.,  2006,
  \mn@doi [\apj] {10.1086/498886}, \href
  {http://adsabs.harvard.edu/abs/2006ApJ...638..963K} {638, 963}

\bibitem[\protect\citeauthoryear{{King}}{{King}}{1988}]{King88}
{King} A.~R.,  1988, \qjras, \href
  {http://adsabs.harvard.edu/abs/1988QJRAS..29....1K} {29, 1}

\bibitem[\protect\citeauthoryear{{Kirsch}, {Mukerjee}, {Breitfellner},
  {Djavidnia}, {Freyberg}, {Kendziorra}  \& {Smith}}{{Kirsch}
  et~al.}{2004}]{Kirsch04}
{Kirsch} M.~G.~F.,  {Mukerjee} K.,  {Breitfellner} M.~G.,  {Djavidnia} S.,
  {Freyberg} M.~J.,  {Kendziorra} E.,   {Smith} M.~J.~S.,  2004, \mn@doi [\aap]
  {10.1051/0004-6361:200400022}, \href
  {http://adsabs.harvard.edu/abs/2004A%26A...423L...9K} {423, L9}

\bibitem[\protect\citeauthoryear{{Lazaridis} et~al.,}{{Lazaridis}
  et~al.}{2011}]{Lazaridis2011a}
{Lazaridis} K.,  et~al., 2011, \mn@doi [\mnras]
  {10.1111/j.1365-2966.2011.18610.x}, \href
  {http://adsabs.harvard.edu/abs/2011MNRAS.414.3134L} {414, 3134}

\bibitem[\protect\citeauthoryear{{Lyne} \& {Graham-Smith}}{{Lyne} \&
  {Graham-Smith}}{1990}]{Lyne90}
{Lyne} A.~G.,  {Graham-Smith} F.,  1990, {Pulsar astronomy}

\bibitem[\protect\citeauthoryear{{Markwardt} \& {Swank}}{{Markwardt} \&
  {Swank}}{2005}]{Markwardt05}
{Markwardt} C.~B.,  {Swank} J.~H.,  2005, The Astronomer's Telegram, \href
  {http://adsabs.harvard.edu/abs/2005ATel..495....1M} {495, 1}

\bibitem[\protect\citeauthoryear{{Muno}, {{\"O}zel}  \& {Chakrabarty}}{{Muno}
  et~al.}{2002}]{Muno02}
{Muno} M.~P.,  {{\"O}zel} F.,   {Chakrabarty} D.,  2002, \mn@doi [\apj]
  {10.1086/344152}, \href {http://adsabs.harvard.edu/abs/2002ApJ...581..550M}
  {581, 550}

\bibitem[\protect\citeauthoryear{{Muno}, {{\"O}zel}  \& {Chakrabarty}}{{Muno}
  et~al.}{2003}]{Muno03}
{Muno} M.~P.,  {{\"O}zel} F.,   {Chakrabarty} D.,  2003, \mn@doi [\apj]
  {10.1086/377447}, \href {http://adsabs.harvard.edu/abs/2003ApJ...595.1066M}
  {595, 1066}

\bibitem[\protect\citeauthoryear{{Ortolani}, {Barbuy}  \& {Bica}}{{Ortolani}
  et~al.}{1994}]{Ortolani94}
{Ortolani} S.,  {Barbuy} B.,   {Bica} E.,  1994, \aaps, \href
  {http://adsabs.harvard.edu/abs/1994A%26AS..108..653O} {108, 653}

\bibitem[\protect\citeauthoryear{{Paczy{\'n}ski}}{{Paczy{\'n}ski}}{1971}]{Paczynski71}
{Paczy{\'n}ski} B.,  1971, \mn@doi [\araa]
  {10.1146/annurev.aa.09.090171.001151}, \href
  {http://adsabs.harvard.edu/abs/1971ARA%26A...9..183P} {9, 183}

\bibitem[\protect\citeauthoryear{{Papitto}, {Menna}, {Burderi}, {Di Salvo},
  {D'Antona}  \& {Robba}}{{Papitto} et~al.}{2005}]{Papitto05}
{Papitto} A.,  {Menna} M.~T.,  {Burderi} L.,  {Di Salvo} T.,  {D'Antona} F.,
  {Robba} N.~R.,  2005, \mn@doi [\apjl] {10.1086/429222}, \href
  {http://adsabs.harvard.edu/abs/2005ApJ...621L.113P} {621, L113}

\bibitem[\protect\citeauthoryear{{Papitto}, {D'A{\`\i}}, {Motta}, {Riggio},
  {Burderi}, {di Salvo}, {Belloni}  \& {Iaria}}{{Papitto}
  et~al.}{2011}]{Papitto11a}
{Papitto} A.,  {D'A{\`\i}} A.,  {Motta} S.,  {Riggio} A.,  {Burderi} L.,  {di
  Salvo} T.,  {Belloni} T.,   {Iaria} R.,  2011, \mn@doi [\aap]
  {10.1051/0004-6361/201015974}, \href
  {http://adsabs.harvard.edu/abs/2011A%26A...526L...3P} {526, L3}

\bibitem[\protect\citeauthoryear{{Papitto}, {de Martino}, {Belloni}, {Burgay},
  {Pellizzoni}, {Possenti}  \& {Torres}}{{Papitto} et~al.}{2015}]{Papitto2015a}
{Papitto} A.,  {de Martino} D.,  {Belloni} T.~M.,  {Burgay} M.,  {Pellizzoni}
  A.,  {Possenti} A.,   {Torres} D.~F.,  2015, \mn@doi [\mnras]
  {10.1093/mnrasl/slv013}, \href
  {http://adsabs.harvard.edu/abs/2015MNRAS.449L..26P} {449, L26}

\bibitem[\protect\citeauthoryear{{Patruno} \& {Watts}}{{Patruno} \&
  {Watts}}{2012}]{Patruno12b}
{Patruno} A.,  {Watts} A.~L.,  2012, preprint, \href
  {http://adsabs.harvard.edu/abs/2012arXiv1206.2727P} {} (\mn@eprint {arXiv}
  {1206.2727})

\bibitem[\protect\citeauthoryear{{Patruno}, {Altamirano}, {Hessels}, {Casella},
  {Wijnands}  \& {van der Klis}}{{Patruno} et~al.}{2009}]{Patruno09a}
{Patruno} A.,  {Altamirano} D.,  {Hessels} J.~W.~T.,  {Casella} P.,  {Wijnands}
  R.,   {van der Klis} M.,  2009, \mn@doi [\apj]
  {10.1088/0004-637X/690/2/1856}, \href
  {http://adsabs.harvard.edu/abs/2009ApJ...690.1856P} {690, 1856}

\bibitem[\protect\citeauthoryear{{Patruno}, {Altamirano}  \&
  {Messenger}}{{Patruno} et~al.}{2010a}]{Patruno10b}
{Patruno} A.,  {Altamirano} D.,   {Messenger} C.,  2010a, \mn@doi [\mnras]
  {10.1111/j.1365-2966.2010.16202.x}, \href
  {http://adsabs.harvard.edu/abs/2010MNRAS.403.1426P} {403, 1426}

\bibitem[\protect\citeauthoryear{{Patruno} et~al.,}{{Patruno}
  et~al.}{2010b}]{Patruno10a}
{Patruno} A.,  et~al., 2010b, The Astronomer's Telegram, \href
  {http://adsabs.harvard.edu/abs/2010ATel.2407....1P} {2407, 1}

\bibitem[\protect\citeauthoryear{{Patruno}, {Bult}, {Gopakumar}, {Hartman},
  {Wijnands}, {van der Klis}  \& {Chakrabarty}}{{Patruno}
  et~al.}{2012}]{Patruno12a}
{Patruno} A.,  {Bult} P.,  {Gopakumar} A.,  {Hartman} J.~M.,  {Wijnands} R.,
  {van der Klis} M.,   {Chakrabarty} D.,  2012, \mn@doi [\apjl]
  {10.1088/2041-8205/746/2/L27}, \href
  {http://adsabs.harvard.edu/abs/2012ApJ...746L..27P} {746, L27}

\bibitem[\protect\citeauthoryear{{Pintore} et~al.,}{{Pintore}
  et~al.}{2014}]{Pintore14}
{Pintore} F.,  et~al., 2014, \mn@doi [\mnras] {10.1093/mnras/stu2001}, \href
  {http://adsabs.harvard.edu/abs/2014MNRAS.445.3745P} {445, 3745}

\bibitem[\protect\citeauthoryear{{Pintore} et~al.,}{{Pintore}
  et~al.}{2016}]{Pintore2016a}
{Pintore} F.,  et~al., 2016, \mn@doi [\mnras] {10.1093/mnras/stw176}, \href
  {http://adsabs.harvard.edu/abs/2016MNRAS.457.2988P} {457, 2988}

\bibitem[\protect\citeauthoryear{{Pletsch} \& {Clark}}{{Pletsch} \&
  {Clark}}{2015}]{Pletsch2015a}
{Pletsch} H.~J.,  {Clark} C.~J.,  2015, \mn@doi [\apj]
  {10.1088/0004-637X/807/1/18}, \href
  {http://adsabs.harvard.edu/abs/2015ApJ...807...18P} {807, 18}

\bibitem[\protect\citeauthoryear{{Pooley} et~al.,}{{Pooley}
  et~al.}{2002}]{Pooley02}
{Pooley} D.,  et~al., 2002, \mn@doi [\apj] {10.1086/340498}, \href
  {http://adsabs.harvard.edu/abs/2002ApJ...573..184P} {573, 184}

\bibitem[\protect\citeauthoryear{{Riggio}, {Di Salvo}, {Burderi}, {Menna},
  {Papitto}, {Iaria}  \& {Lavagetto}}{{Riggio} et~al.}{2008}]{Riggio08}
{Riggio} A.,  {Di Salvo} T.,  {Burderi} L.,  {Menna} M.~T.,  {Papitto} A.,
  {Iaria} R.,   {Lavagetto} G.,  2008, \mn@doi [\apj] {10.1086/533578}, \href
  {http://adsabs.harvard.edu/abs/2008ApJ...678.1273R} {678, 1273}

\bibitem[\protect\citeauthoryear{{Riggio}, {Burderi}, {di Salvo}, {Papitto},
  {D'A{\`\i}}, {Iaria}  \& {Menna}}{{Riggio} et~al.}{2011}]{Riggio11b}
{Riggio} A.,  {Burderi} L.,  {di Salvo} T.,  {Papitto} A.,  {D'A{\`\i}} A.,
  {Iaria} R.,   {Menna} M.~T.,  2011, \mn@doi [\aap]
  {10.1051/0004-6361/201014883}, \href
  {http://adsabs.harvard.edu/abs/2011A%26A...531A.140R} {531, A140}

\bibitem[\protect\citeauthoryear{{Verbunt}}{{Verbunt}}{1993}]{Verbunt93}
{Verbunt} F.,  1993, \mn@doi [\araa] {10.1146/annurev.aa.31.090193.000521},
  \href {http://adsabs.harvard.edu/abs/1993ARA%26A..31...93V} {31, 93}

\bibitem[\protect\citeauthoryear{{Wang}}{{Wang}}{1996}]{Wang96}
{Wang} Y.-M.,  1996, \mn@doi [\apjl] {10.1086/310150}, \href
  {http://adsabs.harvard.edu/abs/1996ApJ...465L.111W} {465, L111}

\bibitem[\protect\citeauthoryear{{in 't Zand} et~al.,}{{in 't Zand}
  et~al.}{1999}]{in-t-Zand99}
{in 't Zand} J.~J.~M.,  et~al., 1999, \aap, \href
  {http://adsabs.harvard.edu/abs/1999A%26A...345..100I} {345, 100}

\bibitem[\protect\citeauthoryear{{in't Zand}, {van Kerkwijk}, {Pooley},
  {Verbunt}, {Wijnands}  \& {Lewin}}{{in't Zand} et~al.}{2001}]{in-t-Zand01}
{in't Zand} J.~J.~M.,  {van Kerkwijk} M.~H.,  {Pooley} D.,  {Verbunt} F.,
  {Wijnands} R.,   {Lewin} W.~H.~G.,  2001, \mn@doi [\apjl] {10.1086/338361},
  \href {http://adsabs.harvard.edu/abs/2001ApJ...563L..41I} {563, L41}

\makeatother
\end{thebibliography}

\label{lastpage}

\end{document}